\documentclass[english,showpacs,preprint,superscriptaddress]{revtex4-1}
\usepackage{lmodern}

\usepackage[T1]{fontenc}
\usepackage{babel}
\usepackage{mathrsfs}
\usepackage{mathtools}
\usepackage{amsmath}
\usepackage{amssymb}
\usepackage[unicode=true,
 bookmarks=false,
 breaklinks=false,pdfborder={0 0 1},backref=false,colorlinks=false]
 {hyperref}

\makeatletter

\usepackage{amsthm}\usepackage{latexsym}\usepackage{bm}\usepackage{amsfonts}

\allowdisplaybreaks
\usepackage{babel}
\usepackage{babel}

\usepackage{babel}

\makeatother

\begin{document}
\title{Discovering exact, gauge-invariant, local energy-momentum conservation
laws for the electromagnetic gyrokinetic system by high-order field
theory on heterogeneous manifolds}
\author{Peifeng Fan}
\email{pffan@szu.edu.cn}

\affiliation{Key Laboratory of Optoelectronic Devices and Systems, College of Physics
and Optoelectronic Engineering, Shenzhen University, Shenzhen 518060,
China}
\affiliation{Advanced Energy Research Center, Shenzhen University, Shenzhen 518060,
China}
\author{Hong Qin}
\email{hongqin@princeton.edu}

\affiliation{Princeton Plasma Physics Laboratory, Princeton University, Princeton,
NJ 08543, USA}
\author{Jianyuan Xiao}
\affiliation{School of Physical Sciences, University of Science and Technology
of China, Hefei, Anhui 230026, China}
\begin{abstract}
Gyrokinetic theory is arguably the most important tool for numerical
studies of transport physics in magnetized plasmas. However, exact
local energy-momentum conservation laws for the electromagnetic gyrokinetic
system have not been found despite continuous effort. Without such
local conservation laws, energy and momentum can be instantaneously
transported across spacetime, which is unphysical and casts doubt
on the validity of numerical simulations based on the gyrokinetic
theory. The standard Noether procedure for deriving conservation laws
from corresponding symmetries does not apply to gyrokinetic systems
because the gyrocenters and electromagnetic field reside on different
manifolds. To overcome this difficulty, we develop a high-order field
theory on heterogeneous manifolds for classical particle-field systems
and apply it to derive exact, local conservation laws, in particular
the energy-momentum conservation laws, for the electromagnetic gyrokinetic
system. A weak Euler-Lagrange equation is established to replace the
standard Euler-Lagrange equation for the particles. It is discovered
that an induced weak Euler-Lagrange current enters the local conservation
laws. And it is the new physics captured by the high-order field theory
on heterogeneous manifolds. A recently developed gauge-symmetrization
method for high-order electromagnetic field theories using the electromagnetic
displacement-potential tensor is applied to render the derived energy-momentum
conservation laws electromagnetic gauge-invariant. 
\end{abstract}
\maketitle

\section{introduction}

Gyrokinetic theory, gradually emerged since the 1960s \citep{Frieman1966,Davidson1967,Catto1978,Antonsen1980,Catto1981,Frieman1982},
has become an indispensable tool for analytical and numerical studies
\citep{Lee1983,Parker1993,Cohen1993,Lin1995,Sydora1996,Dorland2000,Chen2003,Candy2003}
of instabilities and transport in magnetized plasmas, with applications
to magnetic fusion and astrophysics. Modern gyrokinetic theory has
been developed to systematically derive more accurate governing equations.
It began with Littlejohn's treatment of the guiding center dynamics
\citep{Littlejohn1979,Littlejohn1981,Littlejohn1982,Littlejohn1983}
using the Lie perturbation method \citep{Cary1981,Cary1983,Boghosian1987,Littlejohn1984}.
Dubin et al. \citep{Dubin1983} applied the Hamiltonian Lie perturbation
method to derive the gyrokinetic equations for low frequency drift
wave perturbations, followed by Hahm et al. \citep{Hahm1988a,Hahm1988b,Hahm1996}
and Brizard \citep{Brizard1989b,Brizard1989a}. Qin et al. \citep{Qin1998,Qin1999,Qin1999a,Qin2000,Qin2004,Qin2005,Qin2007a}
extended the gyrokinetic model to treat high-frequency dynamics \citep{Qin2000}
and MHD perturbations \citep{Qin1998,Qin1999,Qin1999a}. Sugama introduced
the field theoretical approach for the gyorkinetic models \citep{Sugama2000},
which has been widely adopted since \citep{Brizard2000b,Qin2005,Qin2007a,Brizard2007a,Brizard2016a}.
Present research on gyrokinetic theories focuses on endowing the models
with more physical structures and conservation properties using modern
geometric method \citep{Qin2004,Qin2005,Qin2007a}, with the goal
of achieving improved accuracy \citep{Parra2011,Burby2013,Squire2014,Parra2014}
and fidelity for describing magnetized plasmas. For example, the Euler-Poincare
reduction procedure \citep{Squire2013}, Hamiltonian structure \citep{burby2014hamiltonian,burbythesis}
and explicit gauge independence \citep{burby2019} have been constructed
for gyrokinetic systems. These studies closely couple with the investigation
of structure-preserving geometric algorithms of the guiding center
dynamics \citep{qin2008variational,Qin2009,squire2012gauge,zhang2014canonicalization,ellison2015,burby2017,lelandthesis,kraus2017,ellison2018,Burby2020,Burby2020a,Xiao2021a}
for gyrokinetic simulations with long term accuracy and fidelity.

One conservation property of fundamental importance for theoretical
models in physics is the energy-momentum conservation. The gyrokinetic
theory is no exception. For tokamak physics, the exact energy conservation
law was used to analysis the energy flux and transport property \citep{Stringer1991a}.
The mean flows and radial electric field, crucial for tokamak equilibrium
and stability, are determined by the momentum conservation \citep{Abiteboul2011a,Scott2010}.
Exact conservation laws also serve as tests for the accuracy of numerical
simulations \citep{Garbet2010a,squire2012geometric,xiao2017local,Glasser2020,Xiao2021}.

However, exact local energy-momentum conservation laws for the gyrokinetic
system with fully self-consistent time-dependent electromagnetic field
are still unknown. It is worthwhile to emphasize that we are searching
for local conservation laws instead of the weaker global ones. If
a theoretical model does not admit local energy-momentum conservation
law, energy and momentum can be instantaneously transported across
spacetime, which is unphysical and detrimental for the purpose of
studying energy and momentum transport in magnetized plasmas.

To derive conservation laws, there are two ways to proceed. One can
construct conservation laws by taking various moments of the gyrokinetic
equation system \citep{Dubin1983,Hahm1988a,Brizard1989b}. This approach
is effective for simple systems such as the standard Vlasov-Maxwell
(VM) system in the laboratory phase space, where the moments of energy-momentum
and forms of conservation can be easily guessed based on physical
intuition. However, for more sophisticated systems such as the gyrokinetic
systems, it is difficult to know what moments are involved for the
exact conservation laws.

A better approach is to start from variational principles, or field
theories, and derive conservation laws by identifying first the underpinning
symmetries admitted by the Lagrangians of the systems. This is the
familiar Noether procedure. Low \citep{Low1958} presented the first
variational principle of Vlasov-Maxwell system, where the dynamics
of particles is Lagrangian and that of the electromagnetic field is
Eulerian. Using Low's variational approach for the 6D distribution
function, Sugama et al. \citep{Sugama2013a} derived flux surface
averaged conservation laws of energy and toroidal angular momentum
for a toroidally confined plasma satisfying the Vlasov-Poisson-Ampere
approximation under the Coulomb gauge. 

In principle, such a field theoretical methodology can also be adopted
for gyrokinetic systems or the guiding-center drift kinetic system.
A thorough review of the existing literature shows that the following
work have been done in this regard. i) A local momentum conservation
law for the guiding-center drift kinetic system \citep{Sugama2018}
was derived by Sugama et al. using an Eulerian variational formulation
through the Euler-Poincare reduction procedure \citep{Squire2013,Hirvijoki2020}.
Using the same procedure, a local energy-momentum conservation law
for the guiding-center drift kinetic system was also recently derived
by Hirvijoki et al. \citep{Hirvijoki2020}. ii) Brizard \citep{Brizard2000a}
developed another Eulerian variational principle which requires a
constrained variation of the distribution function on an 8D phase
space. With this formalism, energy and momentum conservation laws
for the guiding-center drift kinetic system \citep{Brizard2016a}
and the gyrokinetic Vlasov-Poisson system \citep{Brizard2011} were
derived, as well as global energy conservation for the electromagnetic
gyrokinetic system \citep{Brizard2010a}. iii) Very recently, Brizard
derived a local energy conservation law for the perturbed electromagnetic
field and distribution function of the electromagnetic gyrokinetic
system when the background field is time-independent \citep{Brizard2017,Brizard2019}. 

Despite these advances, as mentioned above, exact local energy-momentum
conservation laws for the general gyrokinetic Vlasov-Maxwell system
remain elusive. The technical difficulties involved can be viewed
from two different angles. For the Eulerian formalism for gyrokinetic
models, the Euler-Lagrange equation assumes a different form because
the field variations are constrained, and the derivation of conservation
laws from symmetries does not follow the standard Noether procedure
for unconstrained variations. In particular, the well-established
infinitesimal symmetry condition, prolongation and integration by
parts in the jet space \citep{Olver1993} cannot be applied without
modification to constrained variations. Since constrained variations
assume different formats for different applications, there is no established
general formulation for the Noether procedure in the case of constrained
variations. For Low's type of variational principles with mixed Lagrangian
and Eulerian variations, particles (gyrocenters in this case) and
the electromagnetic field reside on different manifolds. The electromagnetic
field is defined on spacetime, but the particles are defined on the
time axis only. This differs from the standard Noether procedure.
These difficulties are not unique to the gyrokinetic theory. They
appear in other systems too. For example, if we choose to derive the
energy-momentum conservation laws for the Vlasov-Maxwell system or
the Vlasov-Poisson system in the laboratory phase space $(\boldsymbol{x},\boldsymbol{v})$
from the corresponding spacetime translation symmetry, we would encounter
exactly the same difficulties. Admittedly, these difficulties are
more prominent for the gyrokinetic system because its Lagrangian depends
on high-order derivatives of the field and the phase space coordinates
for gyrocenters are non-fibrous \citep{Qin2005,Qin2007a}. For the
Vlasov-Maxwell system in the the laboratory phase space $(\boldsymbol{x},\boldsymbol{v})$,
we don't need to go through the symmetry analysis to derive the energy-momentum
conservation, since it can be guessed and proved directly. But to
derive exact conservation laws for gyrokinetic systems, symmetry analysis
seems to be the only viable approach.

Recently, this difficulty is overcome by the development of an alternative
field theory for the classical particle-field system \citep{Qin2014b,Fan2018,Fan2019}.
This new field theory embraces the fact that different components,
i.e., particles and electromagnetic field, reside on heterogeneous
manifolds, and a weak Euler-Lagrange equation was derived to replace
the standard Euler-Lagrange equation for particles. It was shown that
under certain conditions the correspondence between symmetries and
conservation laws is still valid, but with a significant modification.
The weak Euler-Lagrange equation introduces a new current in the corresponding
conservation law. This new current, called weak Euler-Lagrange current,
represents the new physics captured by the field theory on heterogeneous
manifolds \citep{Fan2019}.

The field theory on heterogeneous manifolds has been successfully
applied to find local conservation laws in the Vlasov-Poisson system
and the Vlasov-Darwin system that were previously unknown \citep{Qin2014b,Fan2019}.
In particular, the previous well-known momentum conservation law for
the Vlasov-Darwin system written down by Kaufman and Rostler \citep{Kaufman1971a}
in 1971 without derivation was found to be erroneous, and a correct
momentum conservation was systematically derived using the the field
theory for particle-field system on heterogeneous manifolds \citep{Fan2019}.

In this paper, we extend the field theory for particle-field system
on heterogeneous manifolds to systems with high-order field derivatives
in non-canonical phase space coordinates and apply it to systematically
derive local conservation laws for the electromagnetic gyrokinetic
system from the underpinning spacetime symmetries. In particular,
the exact local energy-momentum conservation laws for the electromagnetic
gyrokinetic system are derived. For gyrokinetic systems, the Finite-Larmor-Radius
(FLR) effect is important, and the Lagrangian density must include
derivatives of the field up to certain desired orders. Therefore,
extending the field theory on heterogeneous manifolds to systems with
high-order field derivatives is a necessary first step. We first extend
the theory to include arbitrary high-order field derivatives, and
then derive the energy-momentum conservation law for the electromagnetic
gyrokinetic system. When the derivatives above the first order are
ignored, the Lagrangian density does not contain any derivatives of
the electromagnetic field $\boldsymbol{E}$ and $\boldsymbol{B}$,
and system reduces to the guiding-center drift kinetic system.

Another difference between the present work and previous studies \citep{Brizard2017,Brizard2019}
is that we don't separate the electromagnetic field into perturbed
and background parts. The field theory and conservation laws are expressed
in terms of the total distribution functions and the 4-potential $\left(\varphi\left(t,\boldsymbol{x}\right),\boldsymbol{A}\left(t,\boldsymbol{x}\right)\right)$.
This ensures that the Lagrangian density does not explicitly depends
on the spacetime coordinates $\boldsymbol{x}$ and $t$, and always
admits exact energy-momentum conservation laws. In previous studies
\citep{Brizard2017,Brizard2019}, the magnetic field are separated
into perturbed and background parts, and conservation laws were derived
for the perturbed fields. However, such conservation laws exist only
when the background field is symmetric with respect to certain spacetime
coordinates. In particular, in the tokamak geometry, the momentum
conservation cannot be established in these previous studies because
the background magnetic field is inhomogeneous.

In the present study, we also adopt a systematic approach to remove
the electromagnetic gauge dependence from the electromagnetic gyrokinetic
system using a gauge-symmetrization method recently developed for
classical charged particle-electromagnetic field theories \citep{Fan2021}.
For field theories involving the electromagnetic field, it is well
known that the Energy-Momentum Tensor (EMT) derived by the Noether
procedure from the underpinning spacetime translation symmetry is
neither gauge invariant (a.k.a. gauge symmetric) nor symmetric with
respect to its tensor indices. The standard Belinfante-Rosenfeld method
\citep{Belinfante1939,Belinfante1940,Rosenfeld1940} symmetrizes the
EMT using a super-potential associated with the angular momentum but
does not necessarily make the EMT gauge invariant for a general field
theory. The result reported in Ref.\,\citep{Fan2021} shows that
a third order tensor called electromagnetic displacement-potential
tensor can be constructed to explicitly remove the gauge dependency
of the EMT for high-order electromagnetic field theories. This method
is applied here to render the exact, local energy-momentum conservation
laws derived for the electromagnetic gyrokinetic system gauge invariant. 

This paper is organized as follows. In Sec.$\thinspace$\ref{sec:general_conservation laws },
we extend the field theory for particle-field systems on heterogeneous
manifolds to systems, such as the gyrokinetic system, with high-order
field derivatives in non-canonical phase space coordinates. The weak
EL equation is developed as necessitated by the fact that classical
particles and fields live on different manifolds. Symmetries for the
systems and the links between the symmetries and conservation laws
are established. In Sec.\,\ref{sec:GK}, the general theory developed
is applied to derive the exact, gauge-invariant, local energy-momentum
conservation laws induced by spacetime translation symmetries for
the electromagnetic gyrokinetic system.

\section{High-order field theory on heterogeneous manifolds \label{sec:general_conservation laws }}

Before specializing to the electromagnetic gyrokinetic system, we
develop a general high-order field theory on heterogeneous manifolds
for particle-field systems using noncanonical phase space coordinates.
A weak Euler-Lagrange equation is derived. Exact local conservation
laws are established from the underpinning symmetries. The weak Euler-Lagrange
current in the conservation laws induced by the weak Euler-Lagrange
equation is the new physics predicted by the field theory on heterogeneous
manifolds.

\subsection{Weak Euler-Lagrangian equation}

We start from the action of particle-field systems and revisit the
field theory on heterogeneous manifolds developed in Refs.$\thinspace$\citep{Qin2014b,Fan2018,Fan2019}.
We extend the theory to include high order field derivatives and use
noncanonical phase space coordinates $\left(\boldsymbol{X}_{a},\boldsymbol{U}_{a}\right)$
for particles. The action of gyrokinetic systems assumes the following
form with the field derivatives up to the $n$-th order,
\begin{equation}
\mathcal{A}=\sum_{a}\int L_{a}\left(t,\boldsymbol{X}_{a},\dot{\boldsymbol{X}}_{a},\boldsymbol{U}_{a},\dot{\boldsymbol{U}}_{a};\mathrm{pr}^{\left(n\right)}\boldsymbol{\psi}\left(t,\boldsymbol{X}_{a}\right)\right)dt+\int\mathcal{L}_{F}\left(t,\boldsymbol{x},\mathrm{pr}^{\left(n\right)}\boldsymbol{\psi}\left(t,\boldsymbol{x}\right)\right)dtd^{3}\boldsymbol{x}.\label{eq:action1}
\end{equation}
In this section, we will work out the field theory for this general
form of action without specializing to gyrokinetic models. The subscript
$a$ labels particles, $\left(\boldsymbol{X}_{a}\left(t\right),\boldsymbol{U}_{a}\left(t\right)\right)$
is the trajectory of the $a$-th particle in phase space over the
time axis. $\boldsymbol{X}_{a}\left(t\right)$ takes value in the
3D laboratory space, and $\boldsymbol{\psi}\left(t,\boldsymbol{x}\right)$
is a vector (or 1-form) field defined on spacetime. For gyrokinetic
system, $\boldsymbol{\psi}$ will be the 4-potentials of the electromagnetic
field , i.e., $\boldsymbol{\psi}=\left(\varphi,\boldsymbol{A}\right)$.
$L_{a}$ is Lagrangian of the $a$-th particle, including the interaction
between the particle and fields. $\mathcal{L}_{F}$ is the Lagrangian
density for the field $\boldsymbol{\psi}$. Here, $\mathrm{pr}^{\left(n\right)}\boldsymbol{\psi}\left(t,\boldsymbol{x}\right)$
as a vector field on the jet space is the prolongation of the field
$\boldsymbol{\psi}\left(t,\boldsymbol{x}\right)$ \citep{Olver1993},
which contains $\boldsymbol{\psi}$ and its derivatives up to the
$n$-th order, i.e.,
\begin{equation}
\mathrm{pr}^{\left(n\right)}\boldsymbol{\psi}\left(t,\boldsymbol{x}\right):=\left(\boldsymbol{\psi},\partial_{\mu_{1}}\boldsymbol{\psi},\cdots,\partial_{\mu_{1}}\partial_{\mu_{2}}\cdots\partial_{\mu_{n}}\boldsymbol{\psi}\right),\label{eq:2}
\end{equation}
where $\partial_{\mu_{i}}\in\left\{ \partial_{t},\partial_{x^{1}},\partial_{x^{2}},\partial_{x^{3}}\right\} ,\text{ }\left(i=1,2,\dots,n\right),$
represents a derivative with respect to one of the spacetime coordinates. 

The difference in the domains of the field and particles is clear
from Eq.$\thinspace$(\ref{eq:action1}). The fields $\boldsymbol{\psi}$
is defined on the 4D spacetime, whereas each particle's trajectory
as a field is just defined on the 1D time axis. The integral of the
Lagrangian density $\mathcal{L}_{F}$ for the field $\boldsymbol{\psi}$
is over spacetime, and the integral of Lagrangian $L_{a}$ for the
$a$-th particle is over the time axis only. Because of this fact,
Noether\textquoteright s procedure of deriving conservation laws from
symmetries is not applicable without modification to the particle-field
system defined by the action $\mathcal{A}$ in Eq.$\thinspace$(\ref{eq:action1}).

To overcome this difficulty, we multiply the first part on the right-hand
side of Eq.$\thinspace$(\ref{eq:action1}) by the identity
\begin{equation}
\int\delta_{a}d^{3}\boldsymbol{x}=1,\label{eq:delta_function}
\end{equation}
where $\delta_{a}\equiv\delta\left(\boldsymbol{x}-\boldsymbol{X}_{a}\left(t\right)\right)$
is Dirac's $\delta$-function. The action $\mathcal{A}$ in Eq.$\thinspace$(\ref{eq:action1})
is then transformed into an integral over spacetime,
\begin{align}
\mathcal{A} & =\int\mathcal{L}dtd^{3}\boldsymbol{x},\mathcal{L}=\sum_{a}\mathcal{L}_{a}+\mathcal{L}_{F},\label{eq:action2}\\
\mathcal{L}_{a} & \left(t,\boldsymbol{x},\boldsymbol{X}_{a},\dot{\boldsymbol{X}}_{a},\boldsymbol{U}_{a},\dot{\boldsymbol{U}}_{a};\mathrm{pr}^{\left(n\right)}\boldsymbol{\psi}\left(t,\boldsymbol{X}_{a}\right)\right)=L_{a}\left(t,\boldsymbol{X}_{a},\dot{\boldsymbol{X}}_{a},\boldsymbol{U}_{a},\dot{\boldsymbol{U}}_{a};\mathrm{pr}^{\left(n\right)}\boldsymbol{\psi}\left(t,\boldsymbol{X}_{a}\right)\right)\delta_{a}.\label{eq:particle_lagrangian_density}
\end{align}

Note that the Lagrangian of the $a$-th particle $L_{a}$ is transformed
to the Lagrangian density $\mathcal{L}_{a}$ by multiplying $\delta_{a}$.
Obviously, the variation of the action we constructed here will not
have any constraints, which will make the variational process easier.
We now calculate how the action given by Eq.$\thinspace$(\ref{eq:action2})
varies in response to the field variations $\delta\boldsymbol{X}_{a},\text{ }\delta\boldsymbol{U}_{a}$
and $\delta\boldsymbol{\psi}$, 
\begin{equation}
\delta\ensuremath{\mathcal{A}}=\sum_{a}\int\left\{ \left[\int\boldsymbol{E}_{\boldsymbol{X}_{a}}\left(\text{\ensuremath{\mathcal{L}}}\right)d^{3}\boldsymbol{x}\right]\cdot\text{\ensuremath{\delta}}\boldsymbol{X}_{a}+\left[\int\boldsymbol{E}_{\boldsymbol{U}_{a}}\left(\text{\ensuremath{\mathcal{L}}}\right)d^{3}\boldsymbol{x}\right]\cdot\text{\ensuremath{\delta}}\boldsymbol{U}_{a}\right\} dt+\int\boldsymbol{E}_{\boldsymbol{\psi}}\left(\text{\ensuremath{\mathcal{L}}}\right)\cdot\delta\boldsymbol{\psi}dtd^{3}\boldsymbol{x},\label{eq:6-1}
\end{equation}
where
\begin{eqnarray}
\boldsymbol{E}_{\boldsymbol{X}_{a}} & \equiv & \frac{\partial}{\partial\boldsymbol{X}_{a}}-\frac{D}{Dt}\frac{\partial}{\partial\dot{\boldsymbol{X}}_{a}},\label{eq:7-1}\\
\boldsymbol{E}_{\boldsymbol{U}_{a}} & \equiv & \frac{\partial}{\partial\boldsymbol{U}_{a}}-\frac{D}{Dt}\frac{\partial}{\partial\dot{\boldsymbol{U}}_{a}},\label{eq:8-1}\\
\boldsymbol{E}_{\boldsymbol{\psi}} & \equiv & \frac{\partial}{\partial\boldsymbol{\psi}}+\sum_{j=1}^{n}\left(-1\right)^{j}D_{\mu_{1}}\cdots D_{\mu_{j}}\frac{\partial}{\partial_{\mu_{1}}\cdots\partial_{\mu_{j}}\boldsymbol{\psi}},\label{eq:9-1}
\end{eqnarray}
are Euler operators with respect to $\boldsymbol{X}_{a},\text{ }\boldsymbol{U}_{a}$
and $\boldsymbol{\psi}$, respectively. In Eq.$\thinspace$(\ref{eq:6-1}),
the terms $\text{\ensuremath{\delta}}\boldsymbol{X}_{a}$ and $\text{\ensuremath{\delta}}\boldsymbol{U}_{a}$
can be taken out from the space integral because they are fields just
defined on the time axis. Applying Hamilton\textquoteright s principle
to Eq.$\thinspace$(\ref{eq:6-1}), we immediately obtain the equations
of motion for particles and fields
\begin{align}
 & \boldsymbol{E}_{\boldsymbol{\psi}}\left(\text{\ensuremath{\mathcal{L}}}\right)=0,\label{eq:10-1}\\
 & \int\boldsymbol{E}_{\boldsymbol{X}_{a}}\left(\text{\ensuremath{\mathcal{L}}}\right)d^{3}\boldsymbol{x}=0,\label{eq:11}\\
 & \int\boldsymbol{E}_{\boldsymbol{U}_{a}}\left(\text{\ensuremath{\mathcal{L}}}\right)d^{3}\boldsymbol{x}=0,\label{eq:12}
\end{align}
by the arbitrariness of $\delta\boldsymbol{X}_{a},\delta\boldsymbol{U}_{a}$
and $\delta\boldsymbol{\psi}$. Equation (\ref{eq:10-1}) is the EL
equation for fields $\boldsymbol{\psi}$. Equations (\ref{eq:11})
and (\ref{eq:12}) are called submanifold Euler-Lagrange equations
for $\boldsymbol{X}_{a}$ and $\boldsymbol{U}_{a}$ because they are
defined only on the time axis after integrating over the spatial dimensions
\citep{Qin2014b,Fan2018,Fan2019}. We can easily prove that the submanifold
EL equations (\ref{eq:11}) and (\ref{eq:12}) are equivalent to the
standard EL equations of $L_{a},$
\begin{equation}
\boldsymbol{E}_{\boldsymbol{X}_{a}}\left(L_{a}\right)=0,\text{ }\boldsymbol{E}_{\boldsymbol{U}_{a}}\left(L_{a}\right)=0,\label{eq:13}
\end{equation}
by substituting the Lagrangian density (\ref{eq:particle_lagrangian_density}).

Our next goal is to derive an explicit expression for $\boldsymbol{E}_{\boldsymbol{U}_{a}}\left(\text{\ensuremath{\mathcal{L}}}\right)$
and $\boldsymbol{E}_{\boldsymbol{X}_{a}}\left(\text{\ensuremath{\mathcal{L}}}\right)$.
From the EL equation (\ref{eq:13}), 
\begin{equation}
\boldsymbol{E}_{\boldsymbol{U}_{a}}\left(\text{\ensuremath{\mathcal{L}}}\right)=\boldsymbol{E}_{\boldsymbol{U}_{a}}\left(L_{a}\right)\delta_{a}=0\label{eq:14}
\end{equation}
because $\delta_{a}$ doesn't depend on $\boldsymbol{U}_{a}$. However,
$\boldsymbol{E}_{\boldsymbol{X}_{a}}\left(\text{\ensuremath{\mathcal{L}}}\right)$
is not zero but a total divergence \citep{Qin2014b,Fan2018,Fan2019},
\begin{align}
\boldsymbol{E}_{\boldsymbol{X}_{a}}\left(\text{\ensuremath{\mathcal{L}}}\right) & =\frac{D}{D\boldsymbol{x}}\cdot\left(\dot{\boldsymbol{X}}_{a}\frac{\partial\mathcal{L}_{a}}{\partial\dot{\boldsymbol{X}}_{a}}-\mathcal{L}_{a}\boldsymbol{I}\right).\label{eq:weak EL}
\end{align}
To prove Eq.\,(\ref{eq:weak EL}), we calculate 
\begin{align*}
\boldsymbol{E}_{\boldsymbol{X}_{a}}\left(\text{\ensuremath{\mathcal{L}}}\right) & =\frac{\partial\left(L_{a}\delta_{a}\right)}{\partial\boldsymbol{X}_{a}}-\frac{D}{Dt}\frac{\partial\left(L_{a}\delta_{a}\right)}{\partial\dot{\boldsymbol{X}}_{a}}\\
 & =\left(\frac{\partial L_{a}}{\partial\boldsymbol{X}_{a}}-\frac{D}{Dt}\frac{\partial L_{a}}{\partial\dot{\boldsymbol{X}}_{a}}\right)\delta_{a}+L_{a}\frac{\partial\delta_{a}}{\partial\boldsymbol{X}_{a}}-\frac{\partial L_{a}}{\partial\dot{\boldsymbol{X}}_{a}}\frac{D\delta_{a}}{Dt}\\
 & =\boldsymbol{E}_{\boldsymbol{X}_{a}}\left(L_{a}\right)\delta_{a}-L_{a}\frac{D\delta_{a}}{D\boldsymbol{x}}+\dot{\boldsymbol{X}}_{a}\cdot\frac{D\delta_{a}}{D\boldsymbol{x}}\frac{\partial L_{a}}{\partial\dot{\boldsymbol{X}}_{a}}\\
 & =\frac{D}{D\boldsymbol{x}}\cdot\left(\dot{\boldsymbol{X}}_{a}\frac{\partial L_{a}}{\partial\dot{\boldsymbol{X}}_{a}}\delta_{a}-L_{a}\delta_{a}\boldsymbol{I}\right)\\
 & =\frac{D}{D\boldsymbol{x}}\cdot\left(\dot{\boldsymbol{X}}_{a}\frac{\partial\mathcal{L}_{a}}{\partial\dot{\boldsymbol{X}}_{a}}-\mathcal{L}_{a}\boldsymbol{I}\right).
\end{align*}
We will refer to Eq.$\thinspace$(\ref{eq:weak EL}) as weak Euler-Lagrange
equation. The qualifier \textquotedblleft weak\textquotedblright{}
here indicates that the spatial integral of $\boldsymbol{E}_{\boldsymbol{X}_{a}}\left(\text{\ensuremath{\mathcal{L}}}\right)$,
instead of $\boldsymbol{E}_{\boldsymbol{X}_{a}}\left(\text{\ensuremath{\mathcal{L}}}\right)$
itself, is zero \citep{Qin2014b,Fan2018,Fan2019}. The weak EL equation
plays a crucial role in connecting symmetries and local conservation
laws for the field theory on heterogeneous manifolds. The non-vanishing
right-hand-side of the weak EL equation (\ref{eq:weak EL}) will induce
a new current in conservation laws \citep{Qin2014b,Fan2018,Fan2019}.
This new current is called the weak Euler-Lagrange current, and it
is the new physics associated with the field theory on heterogeneous
manifolds. 

\subsection{General symmetries and conservation laws}

We now discuss the symmetries and conservation laws. A symmetry of
the action $\mathcal{A}$ is a group of transformations,
\begin{align}
g_{\epsilon}: & \left(t,\boldsymbol{x},\boldsymbol{X}_{a}\left(t\right),\boldsymbol{U}_{a}\left(t\right),\boldsymbol{\psi}\left(t,\boldsymbol{x}\right)\right)\mapsto\left(\tilde{t},\tilde{\boldsymbol{x}},\tilde{\boldsymbol{X}}_{a}\left(\tilde{t}\right),\tilde{\boldsymbol{U}}_{a}\left(\tilde{t}\right),\tilde{\boldsymbol{\psi}}\left(\tilde{t},\tilde{\boldsymbol{x}}\right)\right),\label{eq:16}
\end{align}
such that
\begin{align}
 & \int\mathcal{L}\left(t,\boldsymbol{x},\boldsymbol{X}_{a}\left(t\right),\dot{\boldsymbol{X}}_{a}\left(t\right),\boldsymbol{U}_{a}\left(t\right),\dot{\boldsymbol{U}}_{a}\left(t\right);\mathrm{pr}^{\left(n\right)}\boldsymbol{\psi}\left(t,\boldsymbol{x}\right)\right)dtd^{3}\boldsymbol{x}\nonumber \\
 & =\int\mathcal{L}\left(\tilde{t},\tilde{\boldsymbol{x}},\tilde{\boldsymbol{X}}_{a}\left(\tilde{t}\right),\frac{d\tilde{\boldsymbol{X}}_{a}\left(\tilde{t}\right)}{d\tilde{t}},\tilde{\boldsymbol{U}}_{a}\left(\tilde{t}\right),\frac{d\tilde{\boldsymbol{U}}_{a}\left(\tilde{t}\right)}{d\tilde{t}};\mathrm{pr}^{\left(n\right)}\tilde{\boldsymbol{\psi}}\left(\tilde{t},\tilde{\boldsymbol{x}}\right)\right)d\tilde{t}d\tilde{\boldsymbol{x}}\label{eq:symmetry_condition}
\end{align}
for every subdomain. Here, $g_{\epsilon}$ constitutes a continuous
group of transformations parameterized by $\epsilon$. Equation (\ref{eq:symmetry_condition})
is called symmetry condition. To derive a local conservation law,
an infinitesimal version of the symmetry condition is required. For
this purpose, we take the derivative of Eq.\,(\ref{eq:symmetry_condition})
with respect to $\epsilon$ at $\epsilon=0$,
\begin{equation}
\frac{d}{d\epsilon}|_{0}\int\mathcal{L}\left(\tilde{t},\tilde{\boldsymbol{x}},\tilde{\boldsymbol{X}}_{a}\left(\tilde{t}\right),\frac{d\tilde{\boldsymbol{X}}_{a}\left(\tilde{t}\right)}{d\tilde{t}},\tilde{\boldsymbol{U}}_{a}\left(\tilde{t}\right),\frac{d\tilde{\boldsymbol{U}}_{a}\left(\tilde{t}\right)}{d\tilde{t}};\mathrm{pr}^{\left(n\right)}\tilde{\boldsymbol{\psi}}\left(\tilde{t},\tilde{\boldsymbol{x}}\right)\right)d\tilde{t}d\tilde{\boldsymbol{x}}=0\label{symmetry_condition2}
\end{equation}
Following the procedures in Ref.$\thinspace$\citep{Olver1993}, the
infinitesimal criterion derived from Eq.$\thinspace$ (\ref{symmetry_condition2})
is
\begin{align}
 & \mathrm{pr}^{\left(1,n\right)}\boldsymbol{v}\left(\mathcal{L}\right)+\mathcal{L}\left(\frac{D\xi^{t}}{Dt}+\frac{D}{D\boldsymbol{x}}\cdot\boldsymbol{\xi}\right)=0,\label{eq:infinitesimal_criterion}\\
 & \boldsymbol{v}\coloneqq\frac{d}{d\epsilon}|_{0}g_{\epsilon}\left(t,\boldsymbol{x},\boldsymbol{X}_{a},\boldsymbol{U}_{a},\boldsymbol{\psi}\right)=\xi^{t}\frac{\partial}{\partial t}+\boldsymbol{\xi}\cdot\frac{\partial}{\partial\boldsymbol{x}}+\sum_{a}\boldsymbol{\theta}_{a}\cdot\frac{\partial}{\partial\boldsymbol{X}_{a}}+\sum_{a}\boldsymbol{\zeta}_{a}\cdot\frac{\partial}{\partial\boldsymbol{U}_{a}}+\boldsymbol{\phi}\cdot\frac{\partial}{\partial\boldsymbol{\psi}},\label{eq:20}\\
 & \mathrm{pr}^{\left(1,n\right)}\boldsymbol{v}\coloneqq\frac{d}{d\epsilon}|_{0}\mathrm{pr}^{\left(1,n\right)}g_{\epsilon}\left(t,\boldsymbol{x},\boldsymbol{X}_{a},\boldsymbol{U}_{a},\boldsymbol{\psi}\right)=\frac{d}{d\epsilon}|_{0}\left(\tilde{t},\tilde{\boldsymbol{x}},\tilde{\boldsymbol{X}}_{a},\frac{d\tilde{\boldsymbol{X}}_{a}}{d\tilde{t}},\tilde{\boldsymbol{U}}_{a},\frac{d\tilde{\boldsymbol{U}}_{a}}{d\tilde{t}};\mathrm{pr}^{\left(n\right)}\tilde{\boldsymbol{\psi}}\left(\tilde{t},\tilde{\boldsymbol{x}}\right)\right).\label{eq:21}
\end{align}
Here, $\boldsymbol{v}$ is the infinitesimal generator of the group
of transformations and the vector field $\mathrm{pr}^{\left(1,n\right)}\boldsymbol{v}$
is the prolongation of $\boldsymbol{v}$ defined on the jet space,
which can be explicitly expressed as
\begin{align}
 & \mathrm{pr}^{\left(1,n\right)}\boldsymbol{v}=\boldsymbol{v}+\sum_{a}\boldsymbol{\theta}_{a1}\cdot\frac{\partial}{\partial\dot{\boldsymbol{X}}_{a}}+\sum_{a}\boldsymbol{\zeta}_{a1}\cdot\frac{\partial}{\partial\dot{\boldsymbol{U}}_{a}}+\sum_{j=1}^{n}\phi_{\mu_{1}\cdots\mu_{j}}^{\alpha}\frac{\partial}{\partial\left(\partial_{\mu_{1}}\cdots\partial_{\mu_{j}}\psi^{\alpha}\right)},\label{eq:22}\\
 & \boldsymbol{\theta}_{a1}=\xi^{t}\ddot{\boldsymbol{X}}_{a}+\dot{\boldsymbol{q}}_{a},\text{ }\boldsymbol{\zeta}_{a1}=\xi^{t}\ddot{\boldsymbol{U}}_{a}+\dot{\boldsymbol{p}}_{a},\text{ }\phi_{\mu_{1}\cdots\mu_{j}}^{\alpha}=\xi^{\nu}D_{\mu_{1}}\cdots D_{\mu_{j}}\left(D_{\nu}\psi^{\alpha}\right)+D_{\mu_{1}}\cdots D_{\mu_{j}}Q^{\alpha},\label{eq:23}
\end{align}
where
\begin{align}
 & \boldsymbol{q}_{a}=\boldsymbol{\theta}_{a}-\xi^{t}\dot{\boldsymbol{X}}_{a},\text{ }\boldsymbol{p}_{a}=\boldsymbol{\zeta}_{a}-\xi^{t}\dot{\boldsymbol{U}}_{a},\text{ }Q^{\alpha}=\phi^{\alpha}-\xi^{\nu}D_{\nu}\psi^{\alpha}\label{eq:24}
\end{align}
are the characteristics of the infinitesimal generator $\boldsymbol{v}$.
The superscript $\alpha$ is the index of the fields $\boldsymbol{\phi}$
and \textbf{$\boldsymbol{\psi}$}. The formulations and proofs of
Eqs.$\thinspace$(\ref{eq:22})-(\ref{eq:24}) can be found in Ref.$\thinspace$\citep{Olver1993}.

Having derived the weak EL Eq.$\thinspace$(\ref{eq:weak EL}) and
infinitesimal symmetry criterion (\ref{eq:infinitesimal_criterion}),
we now can establish the conservation law. We cast the infinitesimal
criterion (\ref{eq:infinitesimal_criterion}) into an equivalent form,
\begin{align}
 & \partial_{\nu}\left[\mathcal{L}\xi^{\nu}+\sum_{a}\mathscr{P}_{a}^{\nu}\delta_{a}+\mathbb{P}_{F}^{\nu}\right]+\frac{D}{Dt}\left[\sum_{a}\frac{\partial\mathcal{L}}{\partial\dot{\boldsymbol{X}}_{a}}\cdot\boldsymbol{q}_{a}+\sum_{a}\frac{\partial\mathcal{L}}{\partial\dot{\boldsymbol{U}}_{a}}\cdot\boldsymbol{p}_{a}\right]\nonumber \\
 & +\sum_{a}\left[\boldsymbol{E}_{\boldsymbol{X}_{a}}\left(\mathcal{L}\right)\cdot\boldsymbol{q}_{a}+\boldsymbol{E}_{\boldsymbol{U}_{a}}\left(\mathcal{L}\right)\cdot\boldsymbol{p}_{a}\right]+\boldsymbol{E}_{\boldsymbol{\psi}}\left(\mathcal{L}\right)\cdot\boldsymbol{Q}=0,\label{eq:25}
\end{align}
where the 4-vector fields $\mathscr{P}_{a}^{\nu}$ and $\mathbb{P}_{F}^{v}$
contain high-order derivatives of the field $\boldsymbol{\psi}$.
They are the boundary terms \citep{Noether1918,Olver1993} calculated
by integration by parts,
\begin{equation}
\begin{cases}
\mathscr{P}_{a}^{\nu}=\left(\mathscr{P}_{a}^{0},\mathscr{\boldsymbol{P}}_{a}\right)=\sum_{j=1}^{n}\mathscr{P}_{a(j)}^{\nu},\text{ \ensuremath{\mathscr{P}_{a(j)}^{\nu}}=\ensuremath{\sum_{k=1}^{j}}}\mathscr{P}_{a(j),k}^{\nu},\\
\mathbb{P}_{F}^{\nu}=\left(\mathbb{P}_{F}^{0},\boldsymbol{\mathbb{P}}_{F}\right)=\sum_{j=1}^{n}\mathbb{P}_{F\left(j\right)}^{\nu},\text{ }\mathbb{P}_{F\left(j\right)}^{\nu}=\sum_{k=1}^{j}\mathbb{P}_{F\left(j\right),k}^{\nu}.
\end{cases}\label{eq:26}
\end{equation}
Here, the terms $\mathscr{P}_{a(j),k}^{\nu}$ and $\mathbb{P}_{F\left(j\right),k}^{\nu}$
in Eq.$\thinspace$(\ref{eq:26}) are defined by 
\begin{equation}
\begin{cases}
\mathscr{P}_{a(j),k}^{\nu}=Q^{\alpha}\frac{\partial L_{a}}{\partial\left(\partial_{\nu}\psi^{\alpha}\right)},\\
\mathbb{P}_{F\left(j\right),k}^{\nu}=Q^{\alpha}\frac{\partial\mathcal{L}_{F}}{\partial\left(\partial_{\nu}\psi^{\alpha}\right)},
\end{cases}k=j=1,\label{eq:27}
\end{equation}
\begin{equation}
\begin{cases}
\mathscr{P}_{a(j),k}^{\nu}=\left(-1\right)^{k+1}D_{\mu_{k+1}}\cdots D_{\mu_{j}}Q^{\alpha}\left[\frac{\partial L_{a}}{\partial\left(\partial_{\nu}\partial_{\mu_{k+1}}\cdots\partial_{\mu_{j}}\psi^{\alpha}\right)}\right],\\
\mathbb{P}_{F\left(j\right),k}^{\nu}=\left(-1\right)^{k+1}D_{\mu_{k+1}}\cdots D_{\mu_{j}}Q^{\alpha}\left[\frac{\partial\mathcal{L}_{F}}{\partial\left(\partial_{\nu}\partial_{\mu_{k+1}}\cdots\partial_{\mu_{j}}\psi^{\alpha}\right)}\right],
\end{cases}1=k<j,\label{eq:28}
\end{equation}

\begin{equation}
\begin{cases}
\mathscr{P}_{a(j),k}^{\nu}=\left(-1\right)^{k+1}D_{\mu_{k+1}}\cdots D_{\mu_{j}}Q^{\alpha}\left[D_{\mu_{1}}\cdots D_{\mu_{k-1}}\frac{\partial L_{a}}{\partial\left(\partial_{\mu_{1}}\cdots\partial_{\mu_{k-1}}\partial_{\nu}\partial_{\mu_{k+1}}\cdots\partial_{\mu_{j}}\psi^{\alpha}\right)}\right],\\
\mathbb{P}_{F\left(j\right),k}^{\nu}=\left(-1\right)^{k+1}D_{\mu_{k+1}}\cdots D_{\mu_{j}}Q^{\alpha}\left[D_{\mu_{1}}\cdots D_{\mu_{k-1}}\frac{\partial\mathcal{L}_{F}}{\partial\left(\partial_{\mu_{1}}\cdots\partial_{\mu_{k-1}}\partial_{\nu}\partial_{\mu_{k+1}}\cdots\partial_{\mu_{j}}\psi^{\alpha}\right)}\right],
\end{cases}1<k<j,\label{eq:29}
\end{equation}

\begin{equation}
\begin{cases}
\mathscr{P}_{a(j),k}^{\nu}=\left(-1\right)^{k+1}Q^{\alpha}\left[D_{\mu_{1}}\cdots D_{\mu_{k-1}}\frac{\partial L_{a}}{\partial\left(\partial_{\mu_{1}}\cdots\partial_{\mu_{k-1}}\partial_{\nu}\psi^{\alpha}\right)}\right],\\
\mathbb{P}_{F\left(j\right),k}^{\nu}=\left(-1\right)^{k+1}Q^{\alpha}\left[D_{\mu_{1}}\cdots D_{\mu_{k-1}}\frac{\partial\mathcal{L}_{F}}{\partial\left(\partial_{\mu_{1}}\cdots\partial_{\mu_{k-1}}\partial_{\nu}\psi^{\alpha}\right)}\right],
\end{cases}1<k=j.\label{eq:30-1}
\end{equation}
The last two terms in Eq.$\thinspace$(\ref{eq:25}) vanish due to
the EL equations (\ref{eq:10-1}) and (\ref{eq:14}), while the third
term is not zero because of the weak EL equation (\ref{eq:weak EL})
and induces a new current for system. If the characteristic $\boldsymbol{q}_{a}$
is independent of $\boldsymbol{x}$, the local conservation law of
the symmetry is finally established as
\begin{align}
 & \frac{D}{Dt}\left[\sum_{a}\frac{\partial\mathcal{L}_{a}}{\partial\dot{\boldsymbol{X}}_{a}}\cdot\boldsymbol{q}_{a}+\sum_{a}\frac{\partial\mathcal{L}}{\partial\dot{\boldsymbol{U}}_{a}}\cdot\boldsymbol{p}_{a}+\mathcal{L}\xi^{t}+\sum_{a}\mathscr{P}_{a}^{0}\delta_{a}+\mathbb{P}_{F}^{0}\right]\nonumber \\
 & +\frac{D}{D\boldsymbol{x}}\cdot\left[\mathcal{L}\boldsymbol{\xi}+\sum_{a}\left(\dot{\boldsymbol{X}}_{a}\frac{\partial\mathcal{L}_{a}}{\partial\dot{\boldsymbol{X}}_{a}}-\mathcal{L}_{a}\boldsymbol{I}\right)\cdot\boldsymbol{q}_{a}+\sum_{a}\mathscr{\boldsymbol{P}}_{a}\delta_{a}+\boldsymbol{\mathbb{P}}_{F}\right]=0.\label{eq:general_conservation}
\end{align}
Here, the terms $\dot{\boldsymbol{X}}_{a}$ and $\dot{\boldsymbol{U}}_{a}$
are regarded as functions of $\left(\boldsymbol{X}_{a}\left(t\right),\boldsymbol{U}_{a}\left(t\right)\right)$
through the EL equation (\ref{eq:13}).

\subsection{Statistical form of the conservation laws \label{subsec:Statistical form}}

The local conservation law (\ref{eq:general_conservation}) is written
in terms of particle's phase space coordinates $\left(\boldsymbol{X}_{a}\left(t\right),\boldsymbol{U}_{a}\left(t\right)\right)$
and field $\boldsymbol{\psi}\left(t,\boldsymbol{x}\right)$. To express
it in the statistical form in terms of distribution functions of particles
and field, we classify the particles into several species by their
invariants such as mass and charge. A particle indexed by the subscript
$a$ can be regarded as the $p$-th particle of the $s$-species,
i.e., $a$ is equivalent to a pair of indices,
\begin{equation}
a\sim sp.\label{eq:species classification}
\end{equation}
For each species, the Klimontovich distribution function is defined
to be 
\begin{equation}
F_{s}\left(t,\boldsymbol{x},\boldsymbol{u}\right)\equiv\sum_{p}\left[\delta\left(\boldsymbol{x}-\boldsymbol{X}_{sp}\right)\delta\left(\boldsymbol{u}-\boldsymbol{U}_{sp}\right)\right].\label{eq:Klimontovich_function}
\end{equation}
Functions $L_{a},\text{ }\boldsymbol{q}_{a}$ and $\mathscr{P}_{a}^{\nu}$
in Eq.\,(\ref{eq:general_conservation}) distinguished by the index
$a\sim sp$ are same functions in phase space for the same species.
For such a function $g_{a}\left(\boldsymbol{x},\boldsymbol{u}\right)$,
the label $a\sim sp$ can be replaced just by $s$, i.e.,
\begin{equation}
g_{a}=g_{sp}=g_{s},\label{eq:34}
\end{equation}
In the conservation law (\ref{eq:general_conservation}), the summations
in the form of $\sum_{a}g_{a}\left(\boldsymbol{X}_{a}\left(t\right),\boldsymbol{U}_{a}\left(t\right)\right)\delta_{a}$
can be expressed in terms of the distribution functions $F_{s}\left(t,\boldsymbol{x},\boldsymbol{u}\right),$
\begin{equation}
\sum_{a}g_{a}\left(\boldsymbol{X}_{a}\left(t\right),\boldsymbol{U}_{a}\left(t\right)\right)\delta_{a}=\sum_{s}\int\left[F_{s}\left(t,\boldsymbol{x},\boldsymbol{u}\right)g_{s}\left(\boldsymbol{x},\boldsymbol{u}\right)\right]d^{3}\boldsymbol{u}.\label{eq:35}
\end{equation}

Using Eq.\,(\ref{eq:35}), the conservation law (\ref{eq:general_conservation})
can be equivalently written in the statistical form in terms of the
distribution functions $F_{s}\left(t,\boldsymbol{x},\boldsymbol{u}\right)$
and field $\boldsymbol{\psi}\left(t,\boldsymbol{x}\right)$ as
\begin{align}
 & \frac{D}{Dt}\left[\sum_{s}\int F_{s}\left(\frac{\partial L_{s}}{\partial\dot{\boldsymbol{X}}_{s}}\cdot\boldsymbol{q}_{s}+\frac{\partial L_{s}}{\partial\dot{\boldsymbol{U}}_{s}}\cdot\boldsymbol{p}_{s}+L_{s}\xi^{t}+\mathscr{P}_{s}^{0}\right)d^{3}\boldsymbol{u}+\mathcal{L}_{F}\xi^{t}+\mathbb{P}_{F}^{0}\right]\nonumber \\
 & +\frac{D}{D\boldsymbol{x}}\cdot\left\{ \sum_{s}\int F_{s}\left[\left(\dot{\boldsymbol{X}}_{s}\frac{\partial L_{s}}{\partial\dot{\boldsymbol{X}}_{s}}-L_{s}\boldsymbol{I}\right)\cdot\boldsymbol{q}_{s}+L_{s}\boldsymbol{\xi}+\mathscr{\boldsymbol{P}}_{s}\right]d^{3}\boldsymbol{u}+\mathcal{L}_{F}\boldsymbol{\xi}+\mathbb{\boldsymbol{P}}_{F}\right\} =0,\label{eq:General_Conserva_Klimon}
\end{align}
where $L_{s},\text{ }\boldsymbol{q}_{s},\text{ }\boldsymbol{p}_{s},\text{ }\mathscr{P}_{s}^{\nu},\text{ }\dot{\boldsymbol{X}}_{s},\text{ \ensuremath{\dot{\boldsymbol{U}}_{s}} }$and
$\partial L_{s}/\partial\dot{\boldsymbol{X}}_{s}$ are the functions
in phase space, evaluated at $(t,\boldsymbol{x},\boldsymbol{u}).$

Note that in Eq.\,(\ref{eq:General_Conserva_Klimon}), the index
for individual particles $a$ has been absorbed by the Klimontovich
distribution function $F_{s}\left(t,\boldsymbol{x},\boldsymbol{u}\right),$
which serves as the bridge between particle representation using $\left(\boldsymbol{X}_{a}\left(t\right),\boldsymbol{U}_{a}\left(t\right)\right)$
and distribution function representation. In Sec.\,\ref{sec:GK},
local conservation laws for the electromagnetic gyrokinetic system
will be first established using the particle representation in the
form of Eq.\,(\ref{eq:general_conservation}). They are then transformed
to the statistical form in the form of Eq.\,(\ref{eq:General_Conserva_Klimon})
using this technique.

\section{Exact, gauge-invariant, local energy-momentum conservation laws for
the electromagnetic gyrokinetic system \label{sec:GK}}

In this section, we apply the field theory on heterogeneous manifolds
for particle-field systems developed in Sec.\,\ref{sec:general_conservation laws }
to the electromagnetic gyrokinetic system, and derive the exact, gauge-invariant,
local energy-momentum conservation laws of the system from the underpinning
spacetime translation symmetries. For the general electromagnetic
gyrokinetic system specified by the Lagrangian density in Eq.\,(\ref{eq:General-GK-Lag}),
the final conservation laws are given by Eqs.\,(\ref{eq:energy_conservation-2})
and (\ref{eq:general-GK-momentum-conser}). The derivation is explicitly
illustrated using the first-order system specified by the Lagrangian
density in Eq.\,(\ref{eq:1st-GK-Lag}). 

\subsection{The Electromagnetic gyrokinetic system }

When the field theory on heterogeneous manifolds developed in Sec.\,\ref{sec:general_conservation laws }
is specialized to the electromagnetic gyrokinetic theory, $\boldsymbol{X}_{a}$
is the gyrocenter position, $\boldsymbol{U}_{a}=\left(u_{a},\mu_{a},\theta_{a}\right)$
consists of parallel velocity, magnetic moment and gyrophase, and
the field $\boldsymbol{\psi}\left(t,\boldsymbol{x}\right)=\left(\varphi\left(t,\boldsymbol{x}\right),\boldsymbol{A}\left(t,\boldsymbol{x}\right)\right)$
is the 4-potential. As in the general case, the Lagrangian density
of the system $\mathscr{\mathcal{L}}$ is composed of the field Lagrangian
density $\mathcal{L}_{F}$ and particle Lagrangian $L_{a}$,
\begin{align}
\mathcal{L} & =\mathcal{L}_{F}+\sum_{a}\mathcal{L}_{a},\label{eq:General-GK-Lag}\\
\mathcal{L}_{a} & =L_{a}\delta(\boldsymbol{x}-\boldsymbol{X}_{a}).
\end{align}

For the general electromagnetic gyrokinetic system, $\mathcal{L}_{F}$
is the standard Lagrangian density of the Maxwell field theory,

\begin{equation}
\mathcal{L}_{F}=\frac{1}{8\pi}\left(\boldsymbol{E}^{2}-\boldsymbol{B}^{2}\right),\text{ }\boldsymbol{E}=-\frac{1}{c}\partial_{t}\boldsymbol{A}-\boldsymbol{\nabla}\varphi,\text{ }\boldsymbol{B}=\boldsymbol{\nabla}\times\boldsymbol{A}.\label{eq:46}
\end{equation}
For particles, 
\begin{align}
L_{a} & =L_{0a}+\delta L_{a}=L_{0a}+L_{1a}+......\,,\\
\mathcal{L}_{a} & =L_{a}\delta(\boldsymbol{x}-\boldsymbol{X}_{a})=\mathcal{L}_{0a}+\mathcal{\delta L}_{a}=\mathcal{L}_{0a}+\mathcal{L}_{1a}+......\thinspace,\label{eq:delta-L}
\end{align}
where $L_{0a}$ is the leading order of the Lagrangian $L_{a}$ of
the $a$-th particle, $L_{1a}$ is the first order, etc. And $\delta L_{a}$
represents all high-order terms of of $L_{a}$. The expressions of
$L_{0a}$ and $L_{1a}$ are give by Eqs.\,(\ref{eq:48}) and (\ref{eq:49}),
respectively. The expansion parameter is the small parameter of the
gyrokinetic ordering, i.e., 
\begin{equation}
\epsilon=\text{max}(\rho k,\omega/\Omega)\ll1.\label{eq:g-order}
\end{equation}
Here, $k$ and $\omega$ measure the spacetime scales of the electromagnetic
field $\boldsymbol{E}$ and $\boldsymbol{B}$ associated the total
total 4-potential $(\varphi,\boldsymbol{A})$, and $\rho$ and $\Omega$
are the typical gyro-radius and gyro-frequency of the particles. 

Before carrying out the detailed derivation of the energy-momentum
conservation laws, we shall point out a few features of the electromagnetic
gyrokinetic system defined by Eq.\,(\ref{eq:General-GK-Lag}). In
the gyrokinetic formalism adopted by most researchers, the electromagnetic
potentials (fields) are separated into perturbed and background parts,
\begin{align}
\boldsymbol{A} & \left(t,\boldsymbol{x}\right)=\boldsymbol{A}_{0}\left(t,\boldsymbol{x}\right)+\boldsymbol{A}_{1}\left(t,\boldsymbol{x}\right),\\
\varphi & \left(t,\boldsymbol{x}\right)=\varphi_{0}\left(t,\boldsymbol{x}\right)+\varphi_{1}\left(t,\boldsymbol{x}\right),
\end{align}
where subscript ``$0$'' indicates the background part, and subscript
``$1$'' the perturbed part. Here, $\boldsymbol{A}_{1}\sim\epsilon\boldsymbol{A}_{0}$
and $\varphi_{1}\sim\epsilon\varphi_{0}$. Let $k_{1}$ and $\omega_{1}$
denote the typical wave number and frequency of the electromagnetic
field associated the perturbed 4-potential $(\varphi_{1},\boldsymbol{A}_{1}).$
While gyrokinetic theory requires Eq.\,(\ref{eq:g-order}), it does
allow 
\begin{equation}
\rho k_{1}\sim\omega_{1}/\Omega\sim1.\label{eq:1-order}
\end{equation}
The energy conservation law derived in Refs.\,\citep{Brizard2017,Brizard2019}
is for the perturbed field $(\varphi_{1},\boldsymbol{A}_{1})$ when
the background field $(\varphi_{0},\boldsymbol{A}_{0})$ does not
depend on time explicitly. Because the background magnetic field $\boldsymbol{B}_{0}(\boldsymbol{x})=\nabla\times\boldsymbol{A}_{0}$
depends on $\boldsymbol{x}$, the momentum conservation law in terms
of $(\varphi_{1},\boldsymbol{A}_{1})$ cannot be established in general,
except for the case where $B_{0}(\boldsymbol{x})$ is symmetric with
respect to specific spatial coordinates.

In the present study, we do not separate the electromagnetic potentials
(fields) into perturbed and background parts, and the theory and the
energy-momentum conservation laws are developed for the total field
$(\varphi,\boldsymbol{A})$. Therefore, it is guaranteed that the
Lagrangian density $\mathcal{L}$ defined in Eq.\,(\ref{eq:General-GK-Lag})
does not explicitly depend on the spacetime coordinate $\left(t,\boldsymbol{x}\right)$,
and that the exact local energy-momentum conservation laws always
exist.

It is important to observe that condition (\ref{eq:1-order}) is consistent
with the gyrokinetic ordering (\ref{eq:g-order}), because the amplitude
of the perturbed field is smaller by one order of $\epsilon$. Since
our theory is developed for the total field $(\varphi,\boldsymbol{A})$,
only the gyrokinetic ordering (\ref{eq:g-order}) is required, and
it is valid for cases with condition (\ref{eq:1-order}). To express
the FLR effects of the gyrokinetic systems using the total field $(\varphi,\boldsymbol{A})$,
it is necessary and sufficient to include high-order field derivatives
in the Lagrangian density $\mathcal{L}$, which is the approach we
adopted. The general theory developed include field derivatives to
all orders, and we explicitly work out the first-order theory, which
includes field derivatives up to the second order. 

Without specifying the explicit form of $\mathcal{L}_{F}$ and $L_{a},$
the equations of motion for $\varphi$ and $\boldsymbol{A}$ derived
directly from the Eq.$\thinspace$(\ref{eq:10-1}) are
\begin{align}
E_{\varphi}\left(\mathcal{L}\right) & =\frac{\partial\mathcal{L}}{\partial\varphi}-\frac{D}{D\boldsymbol{x}}\cdot\frac{\partial\mathcal{L}}{\partial\boldsymbol{\nabla}\varphi}\nonumber \\
 & +\sum_{j=1}^{n-1}\left(-1\right)^{j+1}\left(\frac{D}{D\chi^{\mu_{1}}}\cdots\frac{D}{D\chi^{\mu_{j}}}\frac{D}{D\boldsymbol{x}}\right)\cdot\frac{\partial\mathcal{L}}{\partial\left(\partial_{\mu_{1}}\cdots\partial_{\mu_{j}}\boldsymbol{\nabla}\varphi\right)}\nonumber \\
 & =-\frac{D}{D\boldsymbol{x}}\cdot\frac{\partial\mathcal{L}_{F}}{\partial\boldsymbol{\nabla}\varphi}+\frac{\partial}{\partial\varphi}\left(\sum_{a}\mathcal{L}_{a}\right)\nonumber \\
 & +\frac{D}{D\boldsymbol{x}}\cdot\left\{ \sum_{a}\left[-\frac{\partial\mathcal{L}_{a}}{\partial\boldsymbol{\nabla}\varphi}+\sum_{j=1}^{n-1}\left(-1\right)^{j+1}\left(\frac{D}{D\chi^{\mu_{1}}}\cdots\frac{D}{D\chi^{\mu_{j}}}\right)\frac{\partial\mathcal{L}_{a}}{\partial\left(\partial_{\mu_{1}}\cdots\partial_{\mu_{j}}\boldsymbol{\nabla}\varphi\right)}\right]\right\} \nonumber \\
 & =\frac{1}{4\pi}\boldsymbol{\nabla}\cdot\boldsymbol{E}-\rho_{g}+\boldsymbol{\nabla}\cdot\boldsymbol{P}=0,\label{eq:R-4}
\end{align}
\begin{align}
\boldsymbol{E}_{\boldsymbol{A}}\left(\mathcal{L}\right) & =\frac{\partial\mathcal{L}}{\partial\boldsymbol{A}}-\frac{D}{Dt}\frac{\partial\mathcal{L}}{\partial\boldsymbol{A}_{,t}}-\frac{D}{D\boldsymbol{x}}\cdot\frac{\partial\mathcal{L}}{\partial\boldsymbol{\nabla}\boldsymbol{A}}\nonumber \\
 & +\sum_{j=1}^{n-1}\left(-1\right)^{j+1}\left(\frac{D}{D\chi^{\mu_{1}}}\cdots\frac{D}{D\chi^{\mu_{j}}}\frac{D}{Dt}\right)\frac{\partial\mathcal{L}}{\partial\left(\partial_{\mu}\cdots\partial_{\mu_{j}}\boldsymbol{A}_{,t}\right)}\nonumber \\
 & +\sum_{j=1}^{n-1}\left(-1\right)^{j+1}\left(\frac{D}{D\chi^{\mu_{1}}}\cdots\frac{D}{D\chi^{\mu_{j}}}\frac{D}{D\boldsymbol{x}}\right)\cdot\frac{\partial\mathcal{L}}{\partial\left(\partial_{\mu_{1}}\cdots\partial_{\mu_{j}}\boldsymbol{\nabla}\boldsymbol{A}\right)}\nonumber \\
 & =-\frac{D}{Dt}\frac{\partial\mathcal{L}_{F}}{\partial\boldsymbol{A}_{,t}}-\frac{D}{D\boldsymbol{x}}\cdot\frac{\partial\mathcal{L}_{F}}{\partial\boldsymbol{\nabla}\boldsymbol{A}}+\frac{\partial}{\partial\boldsymbol{A}}\left(\sum_{a}\mathcal{L}_{a}\right)\nonumber \\
 & +\frac{D}{Dt}\left\{ \sum_{a}\left[-\frac{\partial\mathcal{L}_{a}}{\partial\boldsymbol{A}_{,t}}+\sum_{j=1}^{n-1}\left(-1\right)^{j+1}\left(\frac{D}{D\chi^{\mu_{1}}}\cdots\frac{D}{D\chi^{\mu_{j}}}\right)\frac{\partial\mathcal{L}_{a}}{\partial\left(\partial_{\mu}\cdots\partial_{\mu_{j}}\boldsymbol{A}_{,t}\right)}\right]\right\} \nonumber \\
 & +\frac{D}{D\boldsymbol{x}}\cdot\left\{ \sum_{a}\left[-\frac{\partial\mathcal{L}_{a}}{\partial\boldsymbol{\nabla}\boldsymbol{A}}+\sum_{j=1}^{n-1}\left(-1\right)^{j+1}\left(\frac{D}{D\chi^{\mu_{1}}}\cdots\frac{D}{D\chi^{\mu_{j}}}\right)\frac{\partial\mathcal{L}_{a}}{\partial\left(\partial_{\mu}\cdots\partial_{\mu_{j}}\boldsymbol{\nabla}\boldsymbol{A}\right)}\right]\right\} ,\nonumber \\
 & =-\frac{1}{4\pi}\left[-\frac{1}{c}\frac{\partial\boldsymbol{E}}{\partial t}+\boldsymbol{\nabla}\times\boldsymbol{B}\right]+\boldsymbol{j}_{g}+\frac{1}{c}\frac{\partial\boldsymbol{P}}{\partial t}+\boldsymbol{\nabla}\times\boldsymbol{M}=0,\label{eq:R-5}
\end{align}
where 
\begin{align}
 & \rho_{g}=-\frac{\partial}{\partial\varphi}\left(\sum_{a}\mathcal{L}_{a}\right),\;\boldsymbol{j}_{g}=\frac{\partial}{\partial\boldsymbol{A}}\left(\sum_{a}\mathcal{L}_{a}\right),\label{eq:R-6}\\
 & \boldsymbol{P}=\sum_{a}\left[\frac{\partial\mathcal{L}_{a}}{\partial\boldsymbol{E}}+\sum_{j=1}^{n-1}\left(-1\right)^{j}D_{\mu_{1}}\cdots D_{\mu_{j}}\frac{\partial\mathcal{L}_{a}}{\partial\left(\partial_{\mu_{1}}\cdots\partial_{\mu_{j}}\boldsymbol{E}\right)}\right],\label{eq:R-7}\\
 & \boldsymbol{M}=\sum_{a}\left[\frac{\partial\mathcal{L}_{a}}{\partial\boldsymbol{B}}+\sum_{j=1}^{n-1}\left(-1\right)^{j}D_{\mu_{1}}\cdots D_{\mu_{j}}\frac{\partial\mathcal{L}_{a}}{\partial\left(\partial_{\mu}\cdots\partial_{\mu_{j}}\boldsymbol{B}\right)}\right].\label{eq:R-8}
\end{align}
The following equations
\begin{align}
 & \frac{\partial\boldsymbol{E}}{\partial\boldsymbol{\nabla}\varphi}=c\frac{\partial\boldsymbol{E}}{\partial\boldsymbol{A}_{,t}}=-\boldsymbol{I},\text{ }\frac{\partial\boldsymbol{B}}{\partial\boldsymbol{\nabla}\boldsymbol{A}}=\frac{\partial}{\partial\boldsymbol{\nabla}\boldsymbol{A}}\left(\boldsymbol{\varepsilon}:\boldsymbol{\nabla}\boldsymbol{A}\right)=\boldsymbol{\varepsilon},\label{eq:42-1}\\
 & \frac{\partial\mathcal{L}_{a}}{\partial\boldsymbol{\nabla}\varphi}=c\frac{\partial\mathcal{L}_{a}}{\partial\boldsymbol{A}_{,t}}=-\frac{\partial\mathcal{L}_{a}}{\partial\boldsymbol{E}},\label{eq:R-9}\\
 & \frac{\partial\mathcal{L}_{a}}{\partial D_{\mu_{1}}\cdots D_{\mu_{j}}\boldsymbol{\nabla}\varphi}=c\frac{\partial\mathcal{L}_{a}}{\partial D_{\mu_{1}}\cdots D_{\mu_{j}}\boldsymbol{A}_{,t}}=-\frac{\partial\mathcal{L}_{a}}{\partial D_{\mu_{1}}\cdots D_{\mu_{j}}\boldsymbol{E}},\text{\ensuremath{\thinspace}}j=1,2,\cdots,n-1,\label{eq:R-10}\\
 & \frac{\partial\mathcal{L}_{a}}{\partial\boldsymbol{\nabla}\boldsymbol{A}}=\boldsymbol{\varepsilon}\cdot\frac{\partial\mathcal{L}_{a}}{\partial\boldsymbol{B}},\frac{\partial\mathcal{L}_{a}}{\partial D_{\mu_{1}}\cdots D_{\mu_{j}}\boldsymbol{\nabla}\boldsymbol{A}}=\boldsymbol{\varepsilon}\cdot\frac{\partial\mathcal{L}_{a}}{\partial D_{\mu_{1}}\cdots D_{\mu_{j}}\boldsymbol{B}},\text{\ensuremath{\thinspace}}j=1,2,\cdots,n-1\label{eq:R-11}
\end{align}
are used in the last steps of Eqs.$\thinspace$(\ref{eq:R-4}) and
(\ref{eq:R-5}), and $\boldsymbol{\varepsilon}$ in Eq.$\thinspace$(\ref{eq:R-11})
is the Levi-Civita symbol in the Cartesian coordinates. In Eq.\,(\ref{eq:R-6}),
$\rho_{g}$ and $\boldsymbol{j}_{g}$ are charge and current densities
of gyrocenter, and $\boldsymbol{P}$ and $\boldsymbol{M}$ in Eqs.$\thinspace$(\ref{eq:R-7})
and (\ref{eq:R-8}) are polarization and magnetization, which contain
field derivatives up to the $n$-th order. Using Eqs.\,(\ref{eq:R-4})
and (\ref{eq:R-5}), the equation of motion for fields $\left(\varphi,\boldsymbol{A}\right)$
are then transformed into 
\begin{align}
 & \boldsymbol{\nabla}\cdot\left(\boldsymbol{E}+4\pi\boldsymbol{P}\right)=4\pi\rho_{g},\label{eq:Maxwell-1}\\
 & \boldsymbol{\nabla}\times\left(\boldsymbol{B}-4\pi\boldsymbol{M}\right)-\frac{1}{c}\frac{\partial}{\partial t}\left(\boldsymbol{E}+4\pi\boldsymbol{P}\right)=4\pi\boldsymbol{j}_{g}.\label{eq:Maxwell-2}
\end{align}

We will derive the exact, gauge-invariant, local energy-momentum conservation
laws for the general electromagnetic gyrokinetic system specified
by the Lagrangian density in Eq.\,(\ref{eq:General-GK-Lag}). The
final conservation laws are given by Eqs.\,(\ref{eq:energy_conservation-2})
and (\ref{eq:general-GK-momentum-conser}). To simplify the presentation,
we only give the detailed derivation for the following first-order
electromagnetic gyrokinetic theory which only keeps $L_{1a}$ in $\delta L_{a}$
\citep{Qin2007a},

\begin{align}
 & \mathcal{L}_{a}=\mathcal{L}_{a0}+\mathcal{L}_{a1}=\left(L_{0a}+L_{1a}\right)\delta(\boldsymbol{x}-\boldsymbol{X}_{a}),\label{eq:1st-GK-Lag}\\
 & L_{0a}=\frac{q_{a}}{c}\boldsymbol{A}_{a}^{\dagger}\cdot\dot{\boldsymbol{X}}_{a}-q_{a}H_{a},\label{eq:48}\\
 & L_{1a}=-\frac{m_{a}c}{q_{a}}\mu_{a}\boldsymbol{R}_{a}\cdot\dot{\boldsymbol{X}}_{a}-\frac{m_{a}c}{q_{a}}\left\{ \left(\boldsymbol{E}_{a\perp}^{\dagger}-\frac{u_{a}}{c}\boldsymbol{B}_{a}^{\dagger}\times\boldsymbol{b}\right)\cdot\frac{\mu_{a}c}{2BB_{a\parallel}^{\dagger}}\boldsymbol{\nabla}B\right.\nonumber \\
 & \left.\vphantom{\frac{w^{2}}{4B_{0}^{2}B_{\parallel}^{\dagger}}_{a}}+\frac{\mu_{a}u_{a}}{2}\boldsymbol{b}\cdot\boldsymbol{\nabla}\times\boldsymbol{b}-\frac{\mu_{a}c}{2B}\left(\boldsymbol{\nabla}\cdot\boldsymbol{E}-\boldsymbol{bb}:\boldsymbol{\nabla}\boldsymbol{E}\right)-\frac{\mu_{a}}{m_{a}}R_{a}^{0}\right\} ,\label{eq:49}\\
 & \boldsymbol{R}_{a}=\left(\boldsymbol{\nabla}\boldsymbol{c}_{a}\right)\cdot\boldsymbol{a}_{a}=\boldsymbol{R}_{a}\left(u_{a},w_{a}\right),\text{ }R_{a}^{0}=-\frac{1}{c}\partial_{t}\boldsymbol{c}_{a}\cdot\boldsymbol{a}_{a}=R_{a}^{0}\left(u_{a},w_{a}\right),\label{eq:50}\\
 & \boldsymbol{c}_{a}=\frac{\boldsymbol{w}_{a}}{w_{a}},\boldsymbol{b}=\frac{\boldsymbol{B}}{B},\text{ }\boldsymbol{a}_{a}=\boldsymbol{b}\times\boldsymbol{c}_{a},\label{eq:51}\\
 & \boldsymbol{E}_{a}^{\dagger}=-\boldsymbol{\nabla}\varphi_{a}^{\dagger}-\frac{1}{c}\partial_{t}\boldsymbol{A}^{\dagger},\text{ }\boldsymbol{B}_{a}^{\dagger}=\boldsymbol{\nabla}\times\boldsymbol{A}_{a}^{\dagger},\label{eq:52}\\
 & \boldsymbol{A}_{a}^{\dagger}=\boldsymbol{A}+\frac{m_{a}c}{q_{a}}u_{a}\boldsymbol{b}+\frac{m_{a}c}{q_{a}}\boldsymbol{D},\text{ }\varphi_{a}^{\dagger}=\varphi+\frac{\mu_{a}}{q_{a}}B,\label{eq:53}\\
 & H_{a}=\frac{1}{2}\frac{m_{a}}{q_{a}}\left(u_{a}^{2}+\boldsymbol{D}^{2}\right)+\frac{\mu_{a}B}{q_{a}}+\varphi,\text{ }\mu_{a}=\frac{m_{a}w_{a}^{2}}{2B},\text{ }\boldsymbol{D}=\frac{c\boldsymbol{E}\times\boldsymbol{B}}{B^{2}}.\label{eq:54}
\end{align}
where $m_{a}$ and $q_{a}$ are mass and charge of the $a$-th particle,
and $\boldsymbol{w}_{a}$ is the perpendicular velocity. The Routh
reduction has been used to decouple the gyrophase dynamics. Note that
the first order Lagrangian $L_{1a}$ contains second-order spacetime
derivatives of the electromagnetic 4-potential $(\varphi,\boldsymbol{A})$.
The prolongation field involved is thus $\mathrm{pr}^{\left(2\right)}\boldsymbol{\psi}\left(t,\boldsymbol{x}\right)$.

From Eqs.$\thinspace$(\ref{eq:R-7}) and (\ref{eq:R-8}), we can
obtain the polarization $\boldsymbol{P}$ and magnetization $\boldsymbol{M}$
for the first-order theory as
\begin{align}
 & \boldsymbol{P}=\boldsymbol{P}_{0}+\boldsymbol{P}_{1},\\
 & \boldsymbol{P}_{0}=\sum_{a}\frac{\partial\mathcal{L}_{0a}}{\partial\boldsymbol{E}}=\sum_{a}\frac{m_{a}c\delta_{a}}{B}\left[\boldsymbol{b}\times\left(\dot{\boldsymbol{X}}_{a}-\boldsymbol{D}\right)\right],\label{eq:polarization}\\
 & \boldsymbol{P}_{1}=\sum_{a}\left[\frac{\partial\mathcal{L}_{1a}}{\partial\boldsymbol{E}}-D_{\mu}\frac{\partial\mathcal{L}_{1a}}{\partial\left(\partial_{\mu}\boldsymbol{E}\right)}\right],\\
 & \boldsymbol{M}=\boldsymbol{M}_{0}+\boldsymbol{M}_{1},\\
 & \boldsymbol{M}_{0}=\sum_{a}\frac{\partial\mathcal{L}_{0a}}{\partial\boldsymbol{B}}\nonumber \\
 & =\sum_{a}\frac{m_{a}c\delta_{a}}{B}\left[\frac{u_{a}}{c}\dot{\boldsymbol{X}}_{a\perp}-\frac{\mu_{a}B}{m_{a}c}\boldsymbol{b}-\frac{\boldsymbol{E}}{B}\times\left(\dot{\boldsymbol{X}}_{a}-\boldsymbol{D}\right)-\frac{2}{c}\left[\left(\dot{\boldsymbol{X}}_{a}-\boldsymbol{D}\right)\cdot\boldsymbol{D}\right]\boldsymbol{b}\right],\label{eq:magnetization}\\
 & \boldsymbol{M}_{1}=\sum_{a}\left[\frac{\partial\mathcal{L}_{1a}}{\partial\boldsymbol{B}}-D_{\mu}\frac{\partial\mathcal{L}_{1a}}{\partial\left(\partial_{\mu}\boldsymbol{B}\right)}\right].
\end{align}
The detailed derivations of Eq.$\thinspace$(\ref{eq:polarization})
and (\ref{eq:magnetization}) are shown in Appendix \ref{sec:polarization and magnetization}. 

\subsection{Time translation symmetry and local energy conservation law \label{sec:times-translation}}

First, we look at the local energy conservation. It is straightforward
to verify that the action for the gyrokinetic system specified by
the Lagrangian density in Eq.\,(\ref{eq:General-GK-Lag}) is invariant
under the time translation, 
\begin{equation}
g_{\epsilon}:\left(t,\boldsymbol{x},\boldsymbol{X}_{a},\boldsymbol{U}_{a},\varphi,\boldsymbol{A}\right)\mapsto\left(\tilde{t},\tilde{\boldsymbol{x}},\tilde{\boldsymbol{X}}_{a},\tilde{\boldsymbol{U}}_{a},\tilde{\varphi},\tilde{\boldsymbol{A}}\right)=\left(t+\epsilon,\boldsymbol{x},\boldsymbol{X}_{a},\boldsymbol{U}_{a},\varphi,\boldsymbol{A}\right),\text{ }\epsilon\in\mathbb{R},
\end{equation}
because the Lagrangian density doesn't contain the time variables
explicitly. Using Eqs.$\thinspace$(\ref{eq:20}) and (\ref{eq:22}),
the infinitesimal generator and its prolongation of the group transformation
are calculated as
\begin{equation}
\boldsymbol{v}=\mathrm{pr}^{\left(1,2\right)}\boldsymbol{v}=\frac{\partial}{\partial t},\label{eq:70}
\end{equation}
where $\xi^{t}=1$, $\boldsymbol{\xi}=0$ and $\boldsymbol{\theta}_{a1}=\phi_{\mu_{1}\cdots\mu_{j}}^{\alpha}=0$
(see Eqs.$\thinspace$(\ref{eq:20})-(\ref{eq:23})). The infinitesimal
criterion (\ref{eq:infinitesimal_criterion}) is reduced to
\begin{equation}
\frac{\partial\mathcal{L}}{\partial t}=0,\label{eq:71}
\end{equation}
which is indeed satisfied as the Lagrangian density doesn't depend
on time explicitly. Because the characteristic of the infinitesimal
generator $\boldsymbol{q}_{a}=\boldsymbol{\theta}_{a}-\xi^{t}\dot{\boldsymbol{X}}_{a}=-\dot{\boldsymbol{X}}_{a}$
is independent of $\boldsymbol{x}$, the infinitesimal criterion (\ref{eq:71})
will induce a conservation law by calculating terms in Eq.$\thinspace$(\ref{eq:general_conservation}).
Using Eqs.$\thinspace$(\ref{eq:24}) and (\ref{eq:26})-(\ref{eq:30-1}),
these terms for the first-order theory specified by Eq.\,(\ref{eq:1st-GK-Lag})
are
\begin{align}
 & \boldsymbol{q}_{a}=-\dot{\boldsymbol{X}}_{a},\text{ }\boldsymbol{p}_{a}=-\dot{\boldsymbol{U}}_{a},\text{ }\boldsymbol{Q}=\left(-\varphi_{,t},-\boldsymbol{A}_{,t}\right),\label{eq:time_tras_characteristic}\\
 & \frac{\partial\mathcal{L}}{\partial\dot{\boldsymbol{X}}_{a}}=\frac{q_{a}}{c}\boldsymbol{A}_{a}^{\dagger}+\frac{\partial\mathcal{L}_{1}}{\partial\dot{\boldsymbol{X}}_{a}},\label{eq:73}\\
 & \sum_{a}\mathscr{P}_{a(1)}^{\nu}\delta_{a}+\mathbb{P}_{F\left(1\right)}^{\nu}=\frac{1}{4\pi}\left(\frac{1}{c}\left(\boldsymbol{E}+4\pi\boldsymbol{P}_{0}\right)\cdot\boldsymbol{A}_{,t},\left(\boldsymbol{E}+4\pi\boldsymbol{P}_{0}\right)\varphi_{,t}+\boldsymbol{A}_{,t}\times\left(\boldsymbol{B}-4\pi\boldsymbol{M}_{0}\right)\right)\nonumber \\
 & +\sum_{a}\mathscr{P}_{1a\left(1\right)}^{\nu}\delta_{a},\label{eq:74}\\
 & \mathscr{P}_{1a\left(1\right)}^{\nu}=\left(\frac{1}{c}\frac{\partial L_{1a}}{\partial\boldsymbol{E}}\cdot\boldsymbol{A}_{,t},\;\frac{\partial L_{1a}}{\partial\boldsymbol{E}}\varphi_{,t}-\boldsymbol{A}_{,t}\times\frac{\partial L_{1a}}{\partial\boldsymbol{B}}\right),\label{eq:75}\\
 & \mathbb{P}_{F\left(2\right)}^{\nu}=0,\label{eq:76}\\
 & \mathscr{P}_{a\left(2\right)}^{\nu}=\left(-\left[\frac{\partial L_{1a}}{\partial\left(\partial_{t}\boldsymbol{E}\right)}\right]\cdot\partial_{t}\boldsymbol{E}-\left[\frac{\partial L_{1a}}{\partial\left(\partial_{t}\boldsymbol{B}\right)}\right]\cdot\partial_{t}\boldsymbol{B}-\frac{1}{c}\left[D_{\mu}\frac{\partial L_{1a}}{\partial\left(\partial_{\mu}\boldsymbol{E}\right)}\right]\cdot\boldsymbol{A}_{,t},\right.\nonumber \\
 & -\left[\frac{\partial L_{1a}}{\partial\left(\boldsymbol{\nabla}\boldsymbol{E}\right)}\right]\cdot\partial_{t}\boldsymbol{E}-\left[\frac{\partial L_{1a}}{\partial\left(\boldsymbol{\nabla}\boldsymbol{B}\right)}\right]\cdot\partial_{t}\boldsymbol{B}-\left[\frac{\partial L_{1a}}{\partial\left(\boldsymbol{\nabla}\boldsymbol{B}\right)}\right]\cdot\partial_{t}\boldsymbol{B}\nonumber \\
 & \left.\vphantom{\left[\frac{\partial\mathcal{L}_{1}}{\partial\left(\partial_{\mu}\right)}\right]}-\left[D_{\mu}\frac{\partial L_{1a}}{\partial\left(\partial_{\mu}\boldsymbol{E}\right)}\right]\varphi_{,t}+\boldsymbol{A}_{,t}\times\left[D_{\mu}\frac{\partial L_{1a}}{\partial\left(\partial_{\mu}\boldsymbol{B}\right)}\right]\right).\label{eq:77}
\end{align}
The detailed derivations of Eqs.$\thinspace$(\ref{eq:74})-(\ref{eq:77})
are shown in Appendix.$\thinspace$\ref{sec:Boundary-terms}. The
velocity $\dot{\boldsymbol{X}}_{a}$, as a function of $(\boldsymbol{X}_{a}(t),\boldsymbol{U}_{a}(t))$,
is determined by the equation of motion of the $a$-th particle \citep{Qin2007a},
which can be obtained by the EL equation (\ref{eq:13}). Substituting
Eqs.$\thinspace$(\ref{eq:time_tras_characteristic})-(\ref{eq:77})
into Eq.$\thinspace$(\ref{eq:general_conservation}), we obtain the
local energy conservation law
\begin{align}
 & \frac{D}{Dt}\left[\sum_{a}q_{a}H_{a}\delta_{a}-\frac{1}{8\pi}\left(\boldsymbol{E}^{2}-\boldsymbol{B}^{2}\right)-\frac{1}{4\pi c}\left(\boldsymbol{E}+4\pi\boldsymbol{P}_{0}\right)\cdot\boldsymbol{A}_{,t}+\sum_{a}\frac{\partial\mathcal{L}_{1}}{\partial\dot{\boldsymbol{X}}_{a}}\cdot\dot{\boldsymbol{X}}_{a}-\mathcal{L}_{1}-\sum_{a}\mathscr{P}_{1a}^{0}\delta_{a}\right]\nonumber \\
 & +\frac{D}{D\boldsymbol{x}}\cdot\left\{ \vphantom{\frac{\partial\mathcal{L}_{1a}}{\partial\dot{\boldsymbol{X}}_{a}}}\sum_{a}q_{a}H_{a}\dot{\boldsymbol{X}}_{a}-\frac{1}{4\pi}\left(\boldsymbol{E}+4\pi\boldsymbol{P}_{0}\right)\varphi_{,t}-\frac{1}{4\pi}\left[\boldsymbol{A}_{,t}\times\left(\boldsymbol{B}-4\pi\boldsymbol{M}_{0}\right)\right]-\sum_{a}\boldsymbol{\mathscr{P}}_{1a}\delta_{a}\right.\nonumber \\
 & +\left.\vphantom{\frac{\partial\mathcal{L}_{1a}}{\partial\dot{\boldsymbol{X}}_{a}}}\sum_{a}\left(\dot{\boldsymbol{X}}_{a}\frac{\partial\mathcal{L}_{1a}}{\partial\dot{\boldsymbol{X}}_{a}}-\mathcal{L}_{1a}\boldsymbol{I}\right)\cdot\dot{\boldsymbol{X}}_{a}\right\} =0,\label{gauge_dep_ener}
\end{align}
where
\begin{align}
 & \mathscr{P}_{1a}^{0}=\mathscr{P}_{1a\left(1\right)}^{0}+\mathscr{P}_{a\left(2\right)}^{0}=\frac{1}{c}\mathfrak{\boldsymbol{p}}_{1a}\cdot\boldsymbol{A}_{,t}-\left[\frac{\partial L_{1a}}{\partial\left(\partial_{t}\boldsymbol{E}\right)}\right]\cdot\partial_{t}\boldsymbol{E}-\left[\frac{\partial L_{1a}}{\partial\left(\partial_{t}\boldsymbol{B}\right)}\right]\cdot\partial_{t}\boldsymbol{B},\label{eq:79}\\
 & \boldsymbol{\mathscr{P}}_{1a}=\boldsymbol{\mathscr{P}}_{1a\left(1\right)}+\boldsymbol{\mathscr{P}}_{a\left(2\right)}\nonumber \\
 & =\varphi_{,t}\mathfrak{\boldsymbol{p}}_{1a}-\boldsymbol{A}_{,t}\times\mathfrak{\boldsymbol{m}}_{1a}-\left[\frac{\partial L_{1a}}{\partial\left(\boldsymbol{\nabla}\boldsymbol{E}\right)}\right]\cdot\partial_{t}\boldsymbol{E}-\left[\frac{\partial L_{1a}}{\partial\left(\boldsymbol{\nabla}\boldsymbol{B}\right)}\right]\cdot\partial_{t}\boldsymbol{B},\label{eq:80}\\
 & \mathfrak{\boldsymbol{p}}_{1a}=\frac{\partial L_{1a}}{\partial\boldsymbol{E}}-\left[D_{\mu}\frac{\partial L_{1a}}{\partial\left(\partial_{\mu}\boldsymbol{E}\right)}\right],\;\mathfrak{\boldsymbol{m}}_{1a}=\frac{\partial L_{1a}}{\partial\boldsymbol{B}}-\left[D_{\mu}\frac{\partial L_{1a}}{\partial\left(\partial_{\mu}\boldsymbol{B}\right)}\right].\label{eq:81}
\end{align}
Here, $\mathfrak{\boldsymbol{p}}_{1a}$ and $\mathfrak{\boldsymbol{m}}_{1a}$
in Eq.$\thinspace$(\ref{eq:81}) are first-order polarization and
magnetization for the $a$-th particle. And $\mathfrak{\boldsymbol{p}}_{1a}$
and $\mathfrak{\boldsymbol{m}}_{1a}$ are obviously guage invariant.

Because electromagnetic field in the field theory is represented by
the 4-potential \textbf{$\left(\varphi,\boldsymbol{A}\right)$}, the
conservation laws depends on gauge explicitly. To remove the explicit
gauge dependency from the Noether procedure, we can add the identity
\begin{equation}
\frac{D}{Dt}\left\{ \frac{D}{D\boldsymbol{x}}\cdot\left[-\left(\frac{\partial\mathcal{L}}{\partial\boldsymbol{E}}-D_{\mu}\frac{\partial\mathcal{L}}{\partial\left(\partial_{\mu}\boldsymbol{E}\right)}\right)\varphi\right]\right\} +\frac{D}{D\boldsymbol{x}}\cdot\left\{ \frac{D}{Dt}\left[\left(\frac{\partial\mathcal{L}}{\partial\boldsymbol{E}}-D_{\mu}\frac{\partial\mathcal{L}}{\partial\left(\partial_{\mu}\boldsymbol{E}\right)}\right)\varphi\right]\right\} =0\label{eq:identity1}
\end{equation}
to Eq.$\thinspace$(\ref{gauge_dep_ener}), and rewrite the two terms
on the left-hand side of Eq.$\thinspace$(\ref{eq:identity1}) as
follows,
\begin{align}
 & \frac{D}{D\boldsymbol{x}}\cdot\left[-\left(\frac{\partial\mathcal{L}}{\partial\boldsymbol{E}}-D_{\mu}\frac{\partial\mathcal{L}}{\partial\left(\partial_{\mu}\boldsymbol{E}\right)}\right)\varphi\right]=\frac{\partial\mathcal{L}_{0}}{\partial\varphi}\varphi+\frac{\partial\mathcal{L}_{0}}{\partial\boldsymbol{E}}\cdot\boldsymbol{\nabla}\varphi-\left(\frac{\partial\mathcal{L}_{1}}{\partial\boldsymbol{E}}-D_{\mu}\frac{\partial\mathcal{L}_{1}}{\partial\left(\partial_{\mu}\boldsymbol{E}\right)}\right)\cdot\boldsymbol{\nabla}\varphi\nonumber \\
 & =-\frac{1}{4\pi}\left(\boldsymbol{E}+4\pi\boldsymbol{P}_{0}\right)\cdot\boldsymbol{\nabla}\varphi-\sum_{a}q_{a}\varphi\delta_{a}-\sum_{a}\left(\mathfrak{\boldsymbol{p}}_{1a}\cdot\boldsymbol{\nabla}\varphi\right)\delta_{a},\label{eq:83}\\
\nonumber \\
 & \frac{D}{Dt}\left[\left(\frac{\partial\mathcal{L}}{\partial\boldsymbol{E}}-D_{\mu}\frac{\partial\mathcal{L}}{\partial\left(\partial_{\mu}\boldsymbol{E}\right)}\right)\varphi\right]=\frac{\partial\mathcal{L}_{0}}{\partial\boldsymbol{E}}\varphi_{,t}+\left(\frac{\partial\mathcal{L}_{1}}{\partial\boldsymbol{E}}-D_{\mu}\frac{\partial\mathcal{L}_{1}}{\partial\left(\partial_{\mu}\boldsymbol{E}\right)}\right)\varphi_{,t}+c\boldsymbol{\nabla}\varphi\times\frac{\partial\mathcal{L}_{0}}{\partial\boldsymbol{B}}\nonumber \\
 & +c\boldsymbol{\nabla}\varphi\times\left[\frac{\partial\mathcal{L}_{1}}{\partial\boldsymbol{B}}-D_{\mu}\frac{\partial\mathcal{L}_{1}}{\partial\left(\partial_{\mu}\boldsymbol{B}\right)}\right]-c\frac{\partial\mathcal{L}_{0}}{\partial\boldsymbol{A}}\varphi-c\boldsymbol{\nabla}\times\left\{ \varphi\left[\frac{\partial\mathcal{L}}{\partial\boldsymbol{B}}-D_{\mu}\frac{\partial\mathcal{L}}{\partial\left(\partial_{\mu}\boldsymbol{B}\right)}\right]\right\} \nonumber \\
 & =\frac{1}{4\pi}\varphi_{,t}\left(\boldsymbol{E}+4\pi\boldsymbol{P}_{0}\right)-\sum_{a}q_{a}\varphi\dot{\boldsymbol{X}}_{a}+\sum_{a}\varphi_{,t}\mathfrak{\boldsymbol{p}}_{1a}\delta_{a}-\frac{c}{4\pi}\boldsymbol{\nabla}\varphi\times\left(\boldsymbol{B}-4\pi\boldsymbol{M}_{0}\right)\nonumber \\
 & +c\boldsymbol{\nabla}\varphi\times\sum_{a}\mathfrak{\boldsymbol{m}}_{1a}\delta_{a}-c\boldsymbol{\nabla}\times\left\{ \varphi\left[\frac{\partial\mathcal{L}}{\partial\boldsymbol{B}}-D_{\mu}\frac{\partial\mathcal{L}}{\partial\left(\partial_{\mu}\boldsymbol{B}\right)}\right]\right\} .\label{eq:84}
\end{align}
The details of the derivation of Eqs.$\thinspace$(\ref{eq:83}) and
(\ref{eq:84}) can be found in Ref.\,\citep{Fan2021}. The resulting
energy conservation is 
\begin{align}
 & \frac{D}{Dt}\left\{ \vphantom{\left[\frac{\partial L_{1a}}{\partial\left(\partial_{t}\boldsymbol{E}\right)}\right]}\sum_{a}\left[\frac{1}{2}m_{a}\left(u_{a}^{2}+\boldsymbol{D}^{2}\right)+\mu_{a}B\right]\delta_{a}+\frac{1}{8\pi}\left(\boldsymbol{E}^{2}+\boldsymbol{B}^{2}\right)+\boldsymbol{P}_{0}\cdot\boldsymbol{E}\right.\nonumber \\
 & \left.\vphantom{\left[\frac{\partial L_{1a}}{\partial\left(\partial_{t}\boldsymbol{E}\right)}\right]}+\sum_{a}\frac{\partial\mathcal{L}_{1}}{\partial\dot{\boldsymbol{X}}_{a}}\cdot\dot{\boldsymbol{X}}_{a}-\mathcal{L}_{1}+\sum_{a}\left[\mathfrak{\boldsymbol{p}}_{1a}\cdot\boldsymbol{E}+\left[\frac{\partial L_{1a}}{\partial\left(\partial_{t}\boldsymbol{E}\right)}\right]\cdot\partial_{t}\boldsymbol{E}+\left[\frac{\partial L_{1a}}{\partial\left(\partial_{t}\boldsymbol{B}\right)}\right]\cdot\partial_{t}\boldsymbol{B}\right]\delta_{a}\right\} \nonumber \\
 & +\frac{D}{D\boldsymbol{x}}\cdot\left\{ \vphantom{\frac{\partial\mathcal{L}_{1a}}{\partial\dot{\boldsymbol{X}}_{a}}}\sum_{a}\left[\frac{1}{2}m_{a}\left(u_{a}^{2}+\boldsymbol{D}^{2}\right)+\mu_{a}B\right]\delta_{a}\dot{\boldsymbol{X}}_{a}+\frac{c}{4\pi}\boldsymbol{E}\times\boldsymbol{B}-c\boldsymbol{E}\times\boldsymbol{M}_{0}\right.\nonumber \\
 & \left.\vphantom{\frac{\partial\mathcal{L}_{1a}}{\partial\dot{\boldsymbol{X}}_{a}}}+\sum_{a}\left[-c\boldsymbol{E}\times\mathfrak{\boldsymbol{m}}_{1a}+\left[\frac{\partial L_{1a}}{\partial\left(\boldsymbol{\nabla}\boldsymbol{E}\right)}\right]\cdot\partial_{t}\boldsymbol{E}+\left[\frac{\partial L_{1a}}{\partial\left(\boldsymbol{\nabla}\boldsymbol{B}\right)}\right]\cdot\partial_{t}\boldsymbol{B}\right]\delta_{a}+\sum_{a}\left(\dot{\boldsymbol{X}}_{a}\frac{\partial\mathcal{L}_{1a}}{\partial\dot{\boldsymbol{X}}_{a}}-\mathcal{L}_{1a}\boldsymbol{I}\right)\cdot\dot{\boldsymbol{X}}_{a}\right\} =0.\label{eq:energy_conservation}
\end{align}
In Eqs.$\thinspace$(\ref{eq:energy_conservation}), $\dot{\boldsymbol{X}}_{a}$
is drift velocity of the guiding center, and it is a function of $\left(\boldsymbol{X}_{a}\left(t\right),\boldsymbol{U}_{a}\left(t\right)\right)$
determined by the EL equation (\ref{eq:13}). The detailed expression
of $\dot{\boldsymbol{X}}_{a}$ can be found in Ref.$\thinspace$\citep{Qin2007a}.

Following the procedure in Sec.\,\ref{subsec:Statistical form},
Eq.$\thinspace$(\ref{eq:energy_conservation}) can be expressed in
terms of the Klimontovich distribution function $F_{s}(t,\boldsymbol{x},\boldsymbol{u})$
and the electromagnetic field,
\begin{align}
 & \frac{D}{Dt}\left\{ \vphantom{\left[\frac{\partial L_{1a}}{\partial\left(\partial_{t}\boldsymbol{E}\right)}\right]}\sum_{s}\int d^{3}\boldsymbol{u}F_{s}\left[\frac{1}{2}m_{s}\left(u_{\parallel}^{2}+\boldsymbol{D}^{2}\right)+\mu B+\boldsymbol{E}\cdot\mathfrak{\boldsymbol{p}}_{0s}\right]+\frac{1}{8\pi}\left(\boldsymbol{E}^{2}+\boldsymbol{B}^{2}\right)\right.\nonumber \\
 & \left.\vphantom{\left[\frac{\partial L_{1a}}{\partial\left(\partial_{t}\boldsymbol{E}\right)}\right]}+\sum_{s}\int d^{3}\boldsymbol{u}F_{s}\left[\frac{\partial L_{1s}}{\partial\dot{\boldsymbol{X}}_{s}}\cdot\dot{\boldsymbol{X}}_{s}-L_{1s}+\mathfrak{\boldsymbol{p}}_{1s}\cdot\boldsymbol{E}+\left[\frac{\partial L_{1s}}{\partial\left(\partial_{t}\boldsymbol{E}\right)}\right]\cdot\partial_{t}\boldsymbol{E}+\left[\frac{\partial L_{1s}}{\partial\left(\partial_{t}\boldsymbol{B}\right)}\right]\cdot\partial_{t}\boldsymbol{B}\right]\right\} \nonumber \\
 & +\frac{D}{D\boldsymbol{x}}\cdot\left\{ \vphantom{\frac{\partial L_{1s}}{\partial\dot{\boldsymbol{X}}_{s}}}\sum_{s}\int d^{3}\boldsymbol{u}F_{s}\left[\frac{1}{2}m_{s}\left(u_{\parallel}^{2}+\boldsymbol{D}^{2}\right)+\mu B-c\boldsymbol{E}\times\boldsymbol{\mathfrak{m}}_{0s}\right]\dot{\boldsymbol{X}}_{s}+\frac{c}{4\pi}\boldsymbol{E}\times\boldsymbol{B}\right.+\sum_{s}\int d^{3}\boldsymbol{u}F_{s}\times\nonumber \\
 & \left.\vphantom{\frac{\partial L_{1s}}{\partial\dot{\boldsymbol{X}}_{s}}}\left[\left(\dot{\boldsymbol{X}}_{s}\frac{\partial L_{1s}}{\partial\dot{\boldsymbol{X}}_{s}}-L_{1s}\boldsymbol{I}\right)\cdot\dot{\boldsymbol{X}}_{s}+\left[-c\boldsymbol{E}\times\mathfrak{\boldsymbol{m}}_{1s}+\left[\frac{\partial L_{1s}}{\partial\left(\boldsymbol{\nabla}\boldsymbol{E}\right)}\right]\cdot\partial_{t}\boldsymbol{E}+\left[\frac{\partial L_{1s}}{\partial\left(\boldsymbol{\nabla}\boldsymbol{B}\right)}\right]\cdot\partial_{t}\boldsymbol{B}\right]\right]\right\} =0,\label{eq:energy_conservation-1}
\end{align}
where 
\begin{align}
 & \boldsymbol{\mathfrak{p}}_{0s}=\frac{\partial L_{0s}}{\partial\boldsymbol{E}}=\sum_{a}\frac{m_{s}c}{B}\left[\boldsymbol{b}\times\left(\dot{\boldsymbol{X}}_{s}-\boldsymbol{D}\right)\right],\label{eq:87}\\
 & \boldsymbol{\mathfrak{m}}_{0s}=\frac{\partial L_{0s}}{\partial\boldsymbol{B}}=\sum_{a}\frac{m_{s}c}{B}\left[\frac{u_{s}}{c}\dot{\boldsymbol{X}}_{a\perp}-\frac{\mu_{s}B}{m_{s}c}\boldsymbol{b}-\frac{\boldsymbol{E}}{B}\times\left(\dot{\boldsymbol{X}}_{s}-\boldsymbol{D}\right)-\frac{2}{c}\left[\left(\dot{\boldsymbol{X}}_{s}-\boldsymbol{D}\right)\cdot\boldsymbol{D}\right]\boldsymbol{b}\right]\label{eq:88}
\end{align}
are the zeroth-order polarization and magnetization for particles
of the s-species. The polarization $\boldsymbol{P}_{1}$ and magnetization
$\boldsymbol{M}_{1}$ are contained in the first-order terms of Eq.$\thinspace$(\ref{eq:energy_conservation-1}).
In the limit of guiding-center drift kinetics, the first-order terms
in Eq.$\thinspace$(\ref{eq:energy_conservation-1}) are neglected,
and we have
\begin{align}
 & \frac{D}{Dt}\left\{ \sum_{s}\int d^{3}\boldsymbol{u}F_{s}\left[\frac{1}{2}m_{s}\left(u_{\parallel}^{2}+\boldsymbol{D}^{2}\right)+\mu B+\boldsymbol{E}\cdot\mathfrak{\boldsymbol{p}}_{0s}\right]+\frac{1}{8\pi}\left(\boldsymbol{E}^{2}+\boldsymbol{B}^{2}\right)\right\} \nonumber \\
 & +\frac{D}{D\boldsymbol{x}}\cdot\left\{ \sum_{s}\int d^{3}\boldsymbol{u}F_{s}\left[\frac{1}{2}m_{s}\left(u_{\parallel}^{2}+\boldsymbol{D}^{2}\right)+\mu B-c\boldsymbol{E}\times\boldsymbol{\mathfrak{m}}_{0s}\right]\dot{\boldsymbol{X}}_{s}+\frac{c}{4\pi}\boldsymbol{E}\times\boldsymbol{B}\right\} =0.\label{eq:Pfirsch_energy}
\end{align}

In the limit of guiding-center drift kinetics, if the $\boldsymbol{E}\times\boldsymbol{B}$
term $\boldsymbol{D}$ in $\mathcal{L}_{a}$ is also ignored, namely,

\begin{equation}
\mathcal{L}_{a}=\left[\left(\frac{q_{a}}{c}\boldsymbol{A}+m_{a}u_{a}\boldsymbol{b}\right)\cdot\dot{\boldsymbol{X}}_{a}-\left(\frac{1}{2}m_{a}u_{a}^{2}+\mu_{a}B+\varphi\right)\right]\delta_{a},\label{eq:90}
\end{equation}
then the polarization vector field $\boldsymbol{P}_{0}$ and magnetization
vector field $\boldsymbol{M}_{0}$ reduce to
\begin{align}
\boldsymbol{P}_{0}=0,\text{ }\boldsymbol{M}_{0}=\sum_{a}\boldsymbol{\mathfrak{m}}_{0a}\delta_{a},\text{ }\boldsymbol{\mathfrak{m}}_{0a}=\frac{m_{a}u_{a}}{B}\dot{\boldsymbol{X}}_{a\perp}-\mu_{a}\boldsymbol{b}.\label{eq:reduced_polarization_magnetization}
\end{align}
Thus, the energy conservation law is further reduced to
\begin{align}
 & \frac{D}{Dt}\left\{ \sum_{a}\left[\frac{1}{2}m_{a}u_{a}^{2}+\mu_{a}B\right]\delta_{a}+\frac{1}{8\pi}\left(\boldsymbol{E}^{2}+\boldsymbol{B}^{2}\right)\right\} \nonumber \\
 & +\frac{D}{D\boldsymbol{x}}\cdot\left\{ \sum_{a}\left[\frac{1}{2}m_{a}u_{a}^{2}+\mu_{a}B\right]\delta_{a}\dot{\boldsymbol{X}}_{a}+\frac{c}{4\pi}\boldsymbol{E}\times\left(\boldsymbol{B}-4\pi\boldsymbol{M}_{0}\right)\right\} =0,\label{eq:drift_energy_conserv}
\end{align}
which, in terms of the distribution function and the electromagnetic
field, is 
\begin{align}
 & \frac{D}{Dt}\left\{ \sum_{s}\int F_{s}\left[\left(\frac{1}{2}m_{s}u_{\parallel}^{2}+\mu B\right)\right]d^{3}\boldsymbol{u}+\frac{1}{8\pi}\left(\boldsymbol{E}^{2}+\boldsymbol{B}^{2}\right)\right\} \nonumber \\
 & +\frac{D}{D\boldsymbol{x}}\cdot\left\{ \sum_{s}\int F_{s}\left[\left(\frac{1}{2}m_{a}u_{\parallel}^{2}+\mu B\right)\dot{\boldsymbol{X}}_{s}-c\boldsymbol{E}\times\boldsymbol{\mathfrak{m}}_{0s}\right]d^{3}\boldsymbol{u}+\frac{c}{4\pi}\boldsymbol{E}\times\boldsymbol{B}\right\} =0.\label{eq:drift_energy_conserv-1}
\end{align}
Equation (\ref{eq:drift_energy_conserv-1}) agrees with the result
of Brizard et al. \citep{Brizard2016a} for guiding-center drift kinetics.
Note that before the present study, local energy conservation law
was not known for the high-order electromagnetic gyrokinetic systems.
Our local energy conservation law for the electromagnetic gyrokinetic
systems (\ref{eq:energy_conservation-1}) and (\ref{eq:energy_conservation-2})
recover the previous known results for the first-order guiding-center
Vlasov-Maxwell system and the drift kinetic system as special cases.

The above derivation of local energy conservation law is for the first-order
theory specified by Eq.\,(\ref{eq:1st-GK-Lag}). For the general
electromagnetic gyrokinetic system of arbitrary high order specified
by Eq.\,(\ref{eq:General-GK-Lag}), an exact, gauge-invariant, local
energy conservation law can be derived using the same method. It is
listed here without detailed derivation, 
\begin{align}
 & \frac{D}{Dt}\left\{ \vphantom{\left[\frac{\partial L_{1a}}{\partial\left(\partial_{t}\boldsymbol{E}\right)}\right]}\sum_{s}\int d^{3}\boldsymbol{u}F_{s}\left[\frac{1}{2}m_{s}\left(u_{\parallel}^{2}+\boldsymbol{D}^{2}\right)+\mu B+\boldsymbol{E}\cdot\mathfrak{\boldsymbol{p}}_{0s}\right]+\frac{1}{8\pi}\left(\boldsymbol{E}^{2}+\boldsymbol{B}^{2}\right)\right.\nonumber \\
 & \left.\vphantom{\left[\frac{\partial L_{1a}}{\partial\left(\partial_{t}\boldsymbol{E}\right)}\right]}+\sum_{s}\int d^{3}\boldsymbol{u}F_{s}\left[\frac{\partial\delta L_{s}}{\partial\dot{\boldsymbol{X}}_{s}}\cdot\dot{\boldsymbol{X}}_{s}-\delta L_{s}+\delta\mathfrak{\boldsymbol{p}}_{s}\cdot\boldsymbol{E}-\delta J_{s}^{0}\right]\right\} \nonumber \\
 & +\frac{D}{D\boldsymbol{x}}\cdot\left\{ \vphantom{\frac{\partial L_{1s}}{\partial\dot{\boldsymbol{X}}_{s}}}\sum_{s}\int d^{3}\boldsymbol{u}F_{s}\left[\frac{1}{2}m_{s}\left(u_{\parallel}^{2}+\boldsymbol{D}^{2}\right)+\mu B-c\boldsymbol{E}\times\boldsymbol{\mathfrak{m}}_{0s}\right]\dot{\boldsymbol{X}}_{s}+\frac{c}{4\pi}\boldsymbol{E}\times\boldsymbol{B}\right.\nonumber \\
 & \left.\vphantom{\frac{\partial L_{1s}}{\partial\dot{\boldsymbol{X}}_{s}}}\sum_{s}\int d^{3}\boldsymbol{u}F_{s}\left[\left(\dot{\boldsymbol{X}}_{s}\frac{\partial\delta L_{s}}{\partial\dot{\boldsymbol{X}}_{s}}-\delta L_{s}\boldsymbol{I}\right)\cdot\dot{\boldsymbol{X}}_{s}+\left[-c\boldsymbol{E}\times\delta\mathfrak{\boldsymbol{m}}_{s}-\delta\boldsymbol{J}_{s}\right]\right]\right\} =0,\label{eq:energy_conservation-2}
\end{align}
where
\begin{align}
\delta\mathfrak{\boldsymbol{p}}_{s} & =\frac{\partial\delta L_{s}}{\partial\boldsymbol{E}}+\sum_{j=1}^{n-1}\left(-1\right)^{j}D_{\mu_{1}}\cdots D_{\mu_{j}}\frac{\partial\delta L_{s}}{\partial\left(\partial_{\mu_{1}}\cdots\partial_{\mu_{j}}\boldsymbol{E}\right)},\label{eq:delta-ps}\\
\delta\mathfrak{\boldsymbol{m}}_{s} & =\frac{\partial\delta L_{s}}{\partial\boldsymbol{B}}+\sum_{j=1}^{n-1}\left(-1\right)^{j}D_{\mu_{1}}\cdots D_{\mu_{j}}\frac{\partial\delta L_{s}}{\partial\left(\partial_{\mu_{1}}\cdots\partial_{\mu_{j}}\boldsymbol{B}\right)},\label{eq:delta-ms}\\
\delta J_{s}^{0} & =\sum_{i=1}^{n}\sum_{j=1}^{i}\left(-1\right)^{j+1}\left[D_{\mu_{1}}\cdots D_{\mu_{j-1}}\frac{\partial\delta L_{s}}{\partial D_{\mu_{1}}\cdots D_{\mu_{j-1}}\partial_{t}D_{\mu_{j+1}}\cdots D_{\mu_{i}}\boldsymbol{E}}\right]\cdot\left(D_{\mu_{j+1}}\cdots D_{\mu_{i}}\partial_{t}\boldsymbol{E}\right)\nonumber \\
 & +\sum_{i=1}^{n}\sum_{j=1}^{i}\left(-1\right)^{j+1}\left[D_{\mu_{1}}\cdots D_{\mu_{j-1}}\frac{\partial\delta L_{s}}{\partial D_{\mu_{1}}\cdots D_{\mu_{j-1}}\partial_{t}D_{\mu_{j+1}}\cdots D_{\mu_{i}}\boldsymbol{B}}\right]\cdot\left(D_{\mu_{j+1}}\cdots D_{\mu_{i}}\partial_{t}\boldsymbol{B}\right),\label{eq:J0s}\\
\delta\boldsymbol{J}_{s} & =\sum_{i=1}^{n}\sum_{j=1}^{i}\left(-1\right)^{j+1}\left[D_{\mu_{1}}\cdots D_{\mu_{j-1}}\frac{\partial\delta L_{s}}{\partial D_{\mu_{1}}\cdots D_{\mu_{j-1}}\boldsymbol{\nabla}D_{\mu_{j+1}}\cdots D_{\mu_{i}}\boldsymbol{E}}\right]\cdot\left(D_{\mu_{j+1}}\cdots D_{\mu_{i}}\partial_{t}\boldsymbol{E}\right)\nonumber \\
 & +\sum_{i=1}^{n}\sum_{j=1}^{i}\left(-1\right)^{j+1}\left[D_{\mu_{1}}\cdots D_{\mu_{j-1}}\frac{\partial\delta L_{s}}{\partial D_{\mu_{1}}\cdots D_{\mu_{j-1}}\boldsymbol{\nabla}D_{\mu_{j+1}}\cdots D_{\mu_{i}}\boldsymbol{B}}\right]\cdot\left(D_{\mu_{j+1}}\cdots D_{\mu_{i}}\partial_{t}\boldsymbol{B}\right).\label{eq:Js}
\end{align}

\subsection{Space translation symmetry and momentum conservation law \label{sec:space-translation}}

We now discuss the space translation symmetry and momentum conservation.
It is straightforward to verify that the action of the gyrokinetic
system specified by Eq.\,(\ref{eq:General-GK-Lag}) is unchanged
under the space translation
\begin{equation}
\left(\tilde{t},\tilde{\boldsymbol{x}},\tilde{\boldsymbol{X}}_{a},\tilde{\boldsymbol{U}}_{a},\tilde{\varphi},\tilde{\boldsymbol{A}}\right)=\left(t,\boldsymbol{x}+\epsilon\boldsymbol{h},\boldsymbol{X}_{a}+\epsilon\boldsymbol{h},\boldsymbol{U}_{a},\varphi,\boldsymbol{A}\right),\label{eq:space_translation}
\end{equation}
where $\boldsymbol{h}$ is an arbitrary constant vector. Note that
this symmetry group transforms both $\boldsymbol{x}$ and $\boldsymbol{X}_{a}$. 

It is worthwhile to emphasize again that in order for the system to
admit spacetime translation symmetry and thus local energy-momentum
conservation laws, we do not separate the electromagnetic field into
background and perturbed components. This is different from other
existing studies in gyrokinetic theory, which separate the background
magnetic field from the perturbed magnetic field, and as a result
no momentum conservation law can be established in these studies for
the plasmas dynamics in tokamaks or devices with inhomogeneous background
magnetic fields. 

The infinitesimal generator corresponding to Eq.\,(\ref{eq:space_translation})
is
\begin{equation}
\boldsymbol{v}=\boldsymbol{h}\cdot\frac{\partial}{\partial\boldsymbol{x}}+\sum_{a}\boldsymbol{h}\cdot\frac{\partial}{\partial\boldsymbol{X}_{a}}.\label{eq:95}
\end{equation}
Because $\xi^{t}=0$, $\boldsymbol{\xi}=\boldsymbol{\theta}_{a}=\boldsymbol{h}$
and $\boldsymbol{\theta}_{a1}=\phi_{\mu_{1}\cdots\mu_{j}}^{\alpha}=0$
(see Eqs.$\thinspace$(\ref{eq:20})-(\ref{eq:23})), the prolongation
of $\boldsymbol{v}$ is the same as $\boldsymbol{v}$, 
\[
\mathrm{pr}^{\left(1,2\right)}\boldsymbol{v}=\boldsymbol{v}.
\]
 The infinitesimal criterion (\ref{eq:infinitesimal_criterion}) is
then satisfied since
\begin{equation}
\boldsymbol{h}\cdot\left(\frac{\partial\mathcal{L}}{\partial\boldsymbol{x}}+\sum_{a}\frac{\partial\mathcal{L}}{\partial\boldsymbol{X}_{a}}\right)=0,\label{eq:96}
\end{equation}
where used is made of the fact that $\partial\delta_{a}/\partial\boldsymbol{x}=-\partial\delta_{a}/\partial\boldsymbol{X}_{a}$.
The characteristics of the infinitesimal generator (\ref{eq:95})
is
\begin{equation}
\boldsymbol{q}_{a}=\boldsymbol{h},\text{ }\boldsymbol{p}_{a}=0,\text{ }\boldsymbol{Q}=-\boldsymbol{h}\cdot\boldsymbol{\nabla}\boldsymbol{\psi}=\left(-\boldsymbol{h}\cdot\boldsymbol{\nabla}\varphi,-\boldsymbol{h}\cdot\boldsymbol{\nabla}\boldsymbol{A}\right).\label{eq:spa_tra_characteristic}
\end{equation}
The infinitesimal criterion (\ref{eq:96}) thus implies a conservation
law because $\boldsymbol{q}_{a}$ is a constant vector field independent
of $\boldsymbol{x}$. 

We now calculate each term in Eq.$\thinspace$(\ref{eq:general_conservation})
for the first-order theory specified by Eq.\,(\ref{eq:1st-GK-Lag})
to obtain the conservation law. Using the definitions of $\mathscr{P}_{a}^{\nu}$
and $\mathbb{P}_{F}^{v}$ (see Eqs.$\thinspace$(\ref{eq:26})-(\ref{eq:30-1})),
the most complicated terms $\sum_{a}\mathscr{P}_{a(1)}^{\nu}\delta_{a}+\mathbb{P}_{F\left(1\right)}^{\nu}$
and $\sum_{a}\mathscr{P}_{a(2)}^{\nu}\delta_{a}+\mathbb{P}_{F\left(2\right)}^{\nu}$
in the conservation law can be explicitly written as
\begin{align}
 & \sum_{a}\mathscr{P}_{a(1)}^{\nu}\delta_{a}+\mathbb{P}_{F\left(1\right)}^{\nu}=\frac{1}{4\pi}\left(\frac{1}{c}\left(\boldsymbol{E}+4\pi\boldsymbol{P}_{0}\right)\cdot\left(\boldsymbol{\nabla}\boldsymbol{A}\right)^{T},\left(\boldsymbol{E}+4\pi\boldsymbol{P}_{0}\right)\boldsymbol{\nabla}\varphi\right.\nonumber \\
 & \left.\vphantom{\left[\left(\boldsymbol{\nabla}\boldsymbol{A}\right)^{T}\right]}-\boldsymbol{\varepsilon}:\left[\left(\boldsymbol{B}-4\pi\boldsymbol{M}_{0}\right)\left(\boldsymbol{\nabla}\boldsymbol{A}\right)^{T}\right]\right)\cdot\boldsymbol{h}+\left(\sum_{a}\boldsymbol{\sigma}_{1a\left(1\right)}^{\nu}\delta_{a}\right)\cdot\boldsymbol{h},\label{eq:98}\\
 & \boldsymbol{\sigma}_{1a\left(1\right)}^{\nu}=\left(\frac{1}{c}\frac{\partial L_{1a}}{\partial\boldsymbol{E}}\cdot\left(\boldsymbol{\nabla}\boldsymbol{A}\right)^{T},\frac{\partial L_{1a}}{\partial\boldsymbol{E}}\boldsymbol{\nabla}\varphi+\frac{\partial L_{1a}}{\partial\boldsymbol{B}}\times\left(\boldsymbol{\nabla}\boldsymbol{A}\right)^{T}\right)\label{eq:99}\\
 & \sum_{a}\mathscr{P}_{a(2)}^{\nu}\delta_{a}+\mathbb{P}_{F\left(2\right)}^{\nu}=\left(\sum_{a}\boldsymbol{\sigma}_{a\left(2\right)}^{\nu}\delta_{a}\right)\cdot\boldsymbol{h},\label{eq:100}\\
 & \boldsymbol{\sigma}_{a\left(2\right)}^{\nu}=\left(-\boldsymbol{\nabla}\boldsymbol{E}\cdot\left[\frac{\partial L_{1a}}{\partial\left(\partial_{t}\boldsymbol{E}\right)}\right]-\boldsymbol{\nabla}\boldsymbol{B}\cdot\left[\frac{\partial L_{1a}}{\partial\left(\partial_{t}\boldsymbol{B}\right)}\right]-\left[\frac{1}{c}D_{\mu}\frac{\partial L_{1a}}{\partial\left(\partial_{\mu}\boldsymbol{E}\right)}\right]\cdot\left(\boldsymbol{\nabla}\boldsymbol{A}\right)^{T},\right.\nonumber \\
 & -\left[\frac{\partial L_{1a}}{\partial\left(\boldsymbol{\nabla}\boldsymbol{E}\right)}\right]\cdot\left(\boldsymbol{\nabla}\boldsymbol{E}\right)^{T}-\left[\frac{\partial L_{1a}}{\partial\left(\boldsymbol{\nabla}\boldsymbol{B}\right)}\right]\cdot\left(\boldsymbol{\nabla}\boldsymbol{B}\right){}^{T}\nonumber \\
 & \left.\vphantom{D_{\mu}\frac{\partial L_{1a}}{\partial\left(\partial_{\mu}\boldsymbol{B}\right)}}-\left[D_{\mu}\frac{\partial L_{1a}}{\partial\left(\partial_{\mu}\boldsymbol{E}\right)}\right]\left(\boldsymbol{\nabla}\varphi\right)-\left[D_{\mu}\frac{\partial L_{1a}}{\partial\left(\partial_{\mu}\boldsymbol{B}\right)}\right]\times\left(\boldsymbol{\nabla}\boldsymbol{A}\right)^{T}\right).\label{eq:101}
\end{align}
The detailed derivations of Eqs.$\thinspace$(\ref{eq:98})-(\ref{eq:101})
are shown in Appendix.$\thinspace$\ref{sec:Boundary-terms}. Substituting
Eqs.$\thinspace$(\ref{eq:98})-(\ref{eq:101}) into Eq.$\thinspace$(\ref{eq:general_conservation}),
we obtain the momentum conservation laws as
\begin{align}
 & \frac{D}{Dt}\left[\sum_{a}\frac{q_{a}}{c}\boldsymbol{A}^{\dagger}\delta_{a}+\frac{1}{4\pi c}\left(\boldsymbol{E}+4\pi\boldsymbol{P}_{0}\right)\cdot\left(\boldsymbol{\nabla}\boldsymbol{A}\right)^{T}+\sum_{a}\frac{\partial\mathcal{L}_{1}}{\partial\dot{\boldsymbol{X}}_{a}}+\sum_{a}\boldsymbol{\sigma}_{a}^{0}\delta_{a}\right]\nonumber \\
 & +\frac{D}{D\boldsymbol{x}}\cdot\left\{ \sum_{a}\frac{q_{a}}{c}\dot{\boldsymbol{X}}_{a}\boldsymbol{A}_{a}^{\dagger}\delta_{a}+\frac{\boldsymbol{E}^{2}-\boldsymbol{B}^{2}}{8\pi}\boldsymbol{I}-\frac{\boldsymbol{B}-4\pi\boldsymbol{M}_{0}}{4\pi}\times\left(\boldsymbol{\nabla}\boldsymbol{A}\right)^{T}\right.\nonumber \\
 & \left.\vphantom{\frac{4\pi\boldsymbol{P}}{4\pi}}+\frac{\boldsymbol{E}+4\pi\boldsymbol{P}_{0}}{4\pi}\boldsymbol{\nabla}\varphi+\sum_{a}\left(\dot{\boldsymbol{X}}_{a}\frac{\partial\mathcal{L}_{1a}}{\partial\dot{\boldsymbol{X}}_{a}}\right)+\sum_{a}\boldsymbol{\sigma}_{a}\delta_{a}\right\} =0,\label{gauge_dep_mom}
\end{align}
where
\begin{align}
 & \boldsymbol{\sigma}_{a}^{0}=\boldsymbol{\sigma}_{1a}^{0}{}_{\left(1\right)}+\boldsymbol{\sigma}_{a\left(2\right)}^{0}=\frac{1}{c}\mathfrak{\boldsymbol{p}}_{1a}\cdot\left(\boldsymbol{\nabla}\boldsymbol{A}\right)^{T}-\boldsymbol{\nabla}\boldsymbol{E}\cdot\left[\frac{\partial L_{1a}}{\partial\left(\partial_{t}\boldsymbol{E}\right)}\right]-\boldsymbol{\nabla}\boldsymbol{B}\cdot\left[\frac{\partial L_{1a}}{\partial\left(\partial_{t}\boldsymbol{B}\right)}\right],\label{eq:103}\\
 & \boldsymbol{\sigma}_{a}=\boldsymbol{\sigma}_{1a\left(1\right)}+\boldsymbol{\sigma}_{a\left(2\right)}\nonumber \\
 & =\mathfrak{\boldsymbol{p}}_{1a}\boldsymbol{\nabla}\varphi+\mathfrak{\boldsymbol{m}}_{1a}\times\left(\boldsymbol{\nabla}\boldsymbol{A}\right)^{T}-\left[\frac{\partial L_{1a}}{\partial\left(\boldsymbol{\nabla}\boldsymbol{E}\right)}\right]\cdot\left(\boldsymbol{\nabla}\boldsymbol{E}\right)^{T}-\left[\frac{\partial L_{1a}}{\partial\left(\boldsymbol{\nabla}\boldsymbol{B}\right)}\right]\cdot\left(\boldsymbol{\nabla}\boldsymbol{B}\right){}^{T}.\label{eq:104}
\end{align}
Akin to the situation of Eq.$\thinspace$(\ref{gauge_dep_ener}) in
Sec.$\thinspace$(\ref{sec:times-translation}), Eq.\,(\ref{gauge_dep_mom})
is gauge dependent. We can add in the following identity
\begin{equation}
\frac{D}{Dt}\left\{ \frac{D}{D\boldsymbol{x}}\cdot\left[-\frac{1}{c}\left(\frac{\partial\mathcal{L}}{\partial\boldsymbol{E}}-D_{\mu}\frac{\partial\mathcal{L}}{\partial\left(\partial_{\mu}\boldsymbol{E}\right)}\right)\boldsymbol{A}\right]\right\} +\frac{D}{D\boldsymbol{x}}\cdot\left\{ \frac{D}{Dt}\left[\frac{1}{c}\left(\frac{\partial\mathcal{L}}{\partial\boldsymbol{E}}-D_{\mu}\frac{\partial\mathcal{L}}{\partial\left(\partial_{\mu}\boldsymbol{E}\right)}\right)\boldsymbol{A}\right]\right\} =0\label{eq:identity2}
\end{equation}
to remove the explicit gauge dependency (see Ref.\,\citep{Fan2021}).
The two terms in Eq.$\thinspace$(\ref{eq:identity2}) can be rewritten
as
\begin{align}
 & \frac{D}{D\boldsymbol{x}}\cdot\left[-\frac{1}{c}\left(\frac{\partial\mathcal{L}}{\partial\boldsymbol{E}}-D_{\mu}\frac{\partial\mathcal{L}}{\partial\left(\partial_{\mu}\boldsymbol{E}\right)}\right)\boldsymbol{A}\right]=-\frac{1}{c}\frac{\partial\mathcal{L}_{0}}{\partial\varphi}\boldsymbol{A}-\frac{1}{c}\left(\frac{\partial\mathcal{L}_{0}}{\partial\boldsymbol{E}}\right)\cdot\boldsymbol{\nabla}\boldsymbol{A}\nonumber \\
 & -\frac{1}{c}\left[\frac{\partial\mathcal{L}_{1}}{\partial\boldsymbol{E}}-D_{\mu}\frac{\partial\mathcal{L}_{1}}{\partial\left(\partial_{\mu}\boldsymbol{E}\right)}\right]\cdot\boldsymbol{\nabla}\boldsymbol{A}=-\sum_{a}\frac{q_{a}}{c}\delta_{a}\boldsymbol{A}-\frac{1}{4\pi c}\left(\boldsymbol{E}+4\pi\boldsymbol{P}_{0}\right)\cdot\boldsymbol{\nabla}\boldsymbol{A}\nonumber \\
 & -\frac{1}{c}\sum_{a}\left(\mathfrak{\boldsymbol{p}}_{1a}\cdot\boldsymbol{\nabla}\boldsymbol{A}\right)\delta_{a},\label{eq:106}\\
\nonumber \\
 & \frac{D}{Dt}\left[\frac{1}{c}\left(\frac{\partial\mathcal{L}}{\partial\boldsymbol{E}}-D_{\mu}\frac{\partial\mathcal{L}}{\partial\left(\partial_{\mu}\boldsymbol{E}\right)}\right)\boldsymbol{A}\right]=-\frac{\partial\mathcal{L}_{0}}{\partial\boldsymbol{A}}\boldsymbol{A}-\left(\frac{\partial\mathcal{L}_{0}}{\partial\boldsymbol{B}}\right)\times\boldsymbol{\nabla}\boldsymbol{A}+\frac{1}{c}\left(\frac{\partial\mathcal{L}_{0}}{\partial\boldsymbol{E}}\right)\boldsymbol{A}_{,t}\nonumber \\
 & -\left[\frac{\partial\mathcal{L}_{1}}{\partial\boldsymbol{B}}-D_{\mu}\frac{\partial\mathcal{L}_{1}}{\partial\left(\partial_{\mu}\boldsymbol{B}\right)}\right]\times\boldsymbol{\nabla}\boldsymbol{A}+\frac{1}{c}\left[\frac{\partial\mathcal{L}_{1}}{\partial\boldsymbol{E}}-D_{\mu}\frac{\partial\mathcal{L}_{1}}{\partial\left(\partial_{\mu}\boldsymbol{E}\right)}\right]\boldsymbol{A}_{,t}+\boldsymbol{\nabla}\times\left[\left(\frac{\partial\mathcal{L}}{\partial\boldsymbol{B}}-D_{\mu}\frac{\partial\mathcal{L}}{\partial\left(\partial_{\mu}\boldsymbol{B}\right)}\right)\boldsymbol{A}\right]\nonumber \\
 & =-\sum_{a}\frac{q_{a}}{c}\delta_{a}\dot{\boldsymbol{X}}_{a}\boldsymbol{A}+\frac{1}{4\pi c}\left(\boldsymbol{E}+4\pi\boldsymbol{P}_{0}\right)\boldsymbol{A}_{,t}+\frac{1}{4\pi}\left(\boldsymbol{B}-4\pi\boldsymbol{M}_{0}\right)\times\boldsymbol{\nabla}\boldsymbol{A}\nonumber \\
 & -\sum_{a}\left(\mathfrak{\boldsymbol{m}}_{1a}\times\boldsymbol{\nabla}\boldsymbol{A}\right)\delta_{a}+\frac{1}{c}\sum_{a}\left(\mathfrak{\boldsymbol{p}}_{1a}\boldsymbol{A}_{,t}\right)\delta_{a}+\boldsymbol{\nabla}\times\left[\left(\frac{\partial\mathcal{L}}{\partial\boldsymbol{B}}-D_{\mu}\frac{\partial\mathcal{L}}{\partial\left(\partial_{\mu}\boldsymbol{B}\right)}\right)\boldsymbol{A}\right].\label{eq:107}
\end{align}
Details of the derivation is shown in Ref.$\thinspace$\citep{Fan2021}.
Substituting Eqs.$\thinspace$(\ref{eq:identity2})-(\ref{eq:107})
into Eq.$\thinspace$(\ref{gauge_dep_mom}), we obtain
\begin{align}
 & \frac{D}{Dt}\left\{ \sum_{a}m_{a}\left(u_{a}\boldsymbol{b}+\boldsymbol{D}\right)\delta_{a}+\frac{\left(\boldsymbol{E}+4\pi\boldsymbol{P}_{0}\right)\times\boldsymbol{B}}{4\pi c}+\sum_{a}\frac{\partial\mathcal{L}_{1}}{\partial\dot{\boldsymbol{X}}_{a}}\vphantom{\left[\frac{\partial L_{1a}}{\partial\left(\partial_{t}\boldsymbol{E}\right)}\right]}\right.\nonumber \\
 & \left.\vphantom{\left[\frac{\partial L_{1a}}{\partial\left(\partial_{t}\boldsymbol{E}\right)}\right]}+\sum_{a}\frac{1}{c}\left(\mathfrak{\boldsymbol{p}}_{1a}\times\boldsymbol{B}\right)\delta_{a}-\left[\frac{\partial L_{1a}}{\partial\left(\partial_{t}\boldsymbol{E}\right)}\right]\cdot\left(\boldsymbol{\nabla}\boldsymbol{E}\right)^{T}\delta_{a}-\left[\frac{\partial L_{1a}}{\partial\left(\partial_{t}\boldsymbol{B}\right)}\right]\cdot\left(\boldsymbol{\nabla}\boldsymbol{B}\right)^{T}\delta_{a}\right\} \nonumber \\
 & +\frac{D}{D\boldsymbol{x}}\cdot\left\{ \sum_{a}\dot{\boldsymbol{X}}_{a}\left(m_{a}u_{a}\boldsymbol{b}+m_{a}\boldsymbol{D}\right)\delta_{a}+\left[\frac{\boldsymbol{E}^{2}+\boldsymbol{B}^{2}}{8\pi}-\left(\boldsymbol{M}_{0}\cdot\boldsymbol{B}\right)\right]\boldsymbol{I}-\frac{1}{4\pi}\boldsymbol{B}\left(\boldsymbol{B}-4\pi\boldsymbol{M}_{0}\right)\right.\nonumber \\
 & -\frac{\left(\boldsymbol{E}+4\pi\boldsymbol{P}_{0}\right)\boldsymbol{E}}{4\pi}+\sum_{a}\left(\dot{\boldsymbol{X}}_{a}\frac{\partial\mathcal{L}_{1a}}{\partial\dot{\boldsymbol{X}}_{a}}\right)-\sum_{a}\mathfrak{\boldsymbol{p}}_{1a}\boldsymbol{E}\delta_{a}+\left[\boldsymbol{B}\mathfrak{\boldsymbol{m}}_{1a}-\left(\boldsymbol{B}\cdot\mathfrak{\boldsymbol{m}}_{1a}\right)\right]\delta_{a}\nonumber \\
 & \left.\vphantom{\left[\frac{\partial L_{1a}}{\partial\left(\boldsymbol{\nabla}\boldsymbol{B}\right)}\right]}-\left[\frac{\partial L_{1a}}{\partial\left(\boldsymbol{\nabla}\boldsymbol{E}\right)}\right]\cdot\left(\boldsymbol{\nabla}\boldsymbol{E}\right)^{T}\delta_{a}-\left[\frac{\partial L_{1a}}{\partial\left(\boldsymbol{\nabla}\boldsymbol{B}\right)}\right]\cdot\left(\boldsymbol{\nabla}\boldsymbol{B}\right){}^{T}\delta_{a}\right\} =0,\label{eq:momentum_conservation}
\end{align}
where used is made of the following equations
\begin{align}
 & \left(\boldsymbol{E}+4\pi\boldsymbol{P}_{0}\right)\cdot\left[\left(\boldsymbol{\nabla}\boldsymbol{A}\right)^{T}-\boldsymbol{\nabla}\boldsymbol{A}\right]=\left(\boldsymbol{E}+4\pi\boldsymbol{P}_{0}\right)\times\boldsymbol{B},\label{eq:109}\\
 & \left(\boldsymbol{B}-4\pi\boldsymbol{M}_{0}\right)\times\left[\boldsymbol{\nabla}\boldsymbol{A}-\left(\boldsymbol{\nabla}\boldsymbol{A}\right)^{T}\right]=\left[\left(\boldsymbol{B}-4\pi\boldsymbol{M}_{0}\right)\cdot\boldsymbol{B}\right]\boldsymbol{I}-\boldsymbol{B}\left(\boldsymbol{B}-4\pi\boldsymbol{M}_{0}\right),\label{eq:110}
\end{align}
Here, the drift velocity $\dot{\boldsymbol{X}}_{a}$ of the guiding
center in Eq.$\thinspace$(\ref{eq:momentum_conservation}) determined
by the EL equation (\ref{eq:13}), which is regarded as a function
of $\left(\boldsymbol{X}_{a}\left(t\right),\boldsymbol{U}_{a}\left(t\right)\right)$.
Using the procedure in Sec.\,\ref{subsec:Statistical form}, the
momentum conservation can be expressed in terms of the the Klimontovich
distribution function $F_{s}(t,\boldsymbol{x},\boldsymbol{u})$ and
the electromagnetic field,

\begin{align}
 & \frac{D}{Dt}\left\{ \sum_{s}\int d^{3}\boldsymbol{u}F_{s}\left[m_{s}\left(u_{\parallel}\boldsymbol{b}+\boldsymbol{D}\right)+\frac{1}{c}\boldsymbol{\mathfrak{p}}_{0s}\times\boldsymbol{B}\right]+\frac{\boldsymbol{E}\times\boldsymbol{B}}{4\pi c}+\sum_{s}\int d^{3}\boldsymbol{u}F_{s}\left[\frac{\partial L_{1s}}{\partial\dot{\boldsymbol{X}}_{s}}+\frac{1}{c}\left(\mathfrak{\boldsymbol{p}}_{1s}\times\boldsymbol{B}\right)\right.\right.\nonumber \\
 & \left.\vphantom{\frac{\partial L_{1s}}{\partial\dot{\boldsymbol{X}}_{s}}}\left.\vphantom{\frac{\partial L_{1s}}{\partial\dot{\boldsymbol{X}}_{s}}}-\boldsymbol{\nabla}\boldsymbol{E}\cdot\left[\frac{\partial L_{1s}}{\partial\left(\partial_{t}\boldsymbol{E}\right)}\right]-\boldsymbol{\nabla}\boldsymbol{B}\cdot\left[\frac{\partial L_{1s}}{\partial\left(\partial_{t}\boldsymbol{B}\right)}\right]\right]\right\} +\frac{D}{D\boldsymbol{x}}\cdot\left\{ \vphantom{\frac{\partial L_{1s}}{\partial\dot{\boldsymbol{X}}_{s}}}\sum_{s}\int d^{3}\boldsymbol{u}F_{s}\left[m_{s}\dot{\boldsymbol{X}}_{s}\left(u_{\parallel}\boldsymbol{b}+\boldsymbol{D}\right)\vphantom{\dot{\boldsymbol{X}}_{s}}\right.\right.\nonumber \\
 & \left.\vphantom{\dot{\boldsymbol{X}}_{s}}+\boldsymbol{B}\mathfrak{\boldsymbol{m}}_{0s}-\left(\boldsymbol{\mathfrak{m}}_{0s}\cdot\boldsymbol{B}\right)\boldsymbol{I}-\mathfrak{\boldsymbol{p}}_{0s}\boldsymbol{E}\right]+\left(\frac{\boldsymbol{E}^{2}+\boldsymbol{B}^{2}}{8\pi}\right)\boldsymbol{I}-\frac{\boldsymbol{E}\boldsymbol{E}+\boldsymbol{B}\boldsymbol{B}}{4\pi}+\sum_{s}\int d^{3}\boldsymbol{u}F_{s}\times\nonumber \\
 & \left.\vphantom{\frac{\partial L_{1s}}{\partial\dot{\boldsymbol{X}}_{s}}}\left[\dot{\boldsymbol{X}}_{s}\frac{\partial L_{1s}}{\partial\dot{\boldsymbol{X}}_{S}}-\mathfrak{\boldsymbol{p}}_{1s}\boldsymbol{E}+\boldsymbol{B}\mathfrak{\boldsymbol{m}}_{1s}-\left(\boldsymbol{B}\cdot\mathfrak{\boldsymbol{m}}_{1s}\right)-\left[\frac{\partial L_{1a}}{\partial\left(\boldsymbol{\nabla}\boldsymbol{E}\right)}\right]\cdot\left(\boldsymbol{\nabla}\boldsymbol{E}\right)^{T}-\left[\frac{\partial L_{1a}}{\partial\left(\boldsymbol{\nabla}\boldsymbol{B}\right)}\right]\cdot\left(\boldsymbol{\nabla}\boldsymbol{B}\right){}^{T}\right]\right\} =0.\label{eq:momentum_conservation-1}
\end{align}

For the special case of guiding-center drift kinetics, the first-order
Lagrangian density $\mathcal{L}_{1a}$ is neglected, and we have
\begin{align}
 & \frac{D}{Dt}\left\{ \sum_{s}\int d^{3}\boldsymbol{u}F_{s}\left[\left(m_{s}u_{\parallel}\boldsymbol{b}+m_{s}\boldsymbol{D}\right)+\frac{1}{c}\boldsymbol{\mathfrak{p}}_{0s}\times\boldsymbol{B}\right]+\frac{\boldsymbol{E}\times\boldsymbol{B}}{4\pi c}\right\} +\frac{D}{D\boldsymbol{x}}\cdot\left\{ \sum_{s}\int d^{3}\boldsymbol{u}F_{s}\left[m_{s}\dot{\boldsymbol{X}}_{s}\left(u_{\parallel}\boldsymbol{b}+\boldsymbol{D}\right)\right.\right.\nonumber \\
\nonumber \\
 & \left.\vphantom{\frac{\partial L_{1s}}{\partial\dot{\boldsymbol{X}}_{s}}}\left.\vphantom{\sum_{s}\int d^{3}\boldsymbol{u}}+\boldsymbol{B}\mathfrak{\boldsymbol{m}}_{0s}-\left(\boldsymbol{\mathfrak{m}}_{0s}\cdot\boldsymbol{B}\right)\boldsymbol{I}-\mathfrak{\boldsymbol{p}}_{0s}\boldsymbol{E}\right]+\left(\frac{\boldsymbol{E}^{2}+\boldsymbol{B}^{2}}{8\pi}\right)\boldsymbol{I}-\frac{\boldsymbol{E}\boldsymbol{E}+\boldsymbol{B}\boldsymbol{B}}{4\pi}\right\} =0.\label{eq:Pfirsch_Momen}
\end{align}

In the limit of guiding-center drift kinetics, if the $\boldsymbol{E}\times\boldsymbol{B}$
term $\boldsymbol{D}$ in $\mathcal{L}_{a}$ is also ignored (see
Eq.$\thinspace$(\ref{eq:90})), then the momentum conservation is
further reduced to
\begin{align}
 & \frac{D}{Dt}\left\{ \sum_{a}m_{a}u_{a}\boldsymbol{b}\delta_{a}+\frac{\boldsymbol{E}\times\boldsymbol{B}}{4\pi c}\right\} +\frac{D}{D\boldsymbol{x}}\cdot\left\{ \sum_{a}m_{a}u_{a}\dot{\boldsymbol{X}}_{a}\boldsymbol{b}\delta_{a}\right.\nonumber \\
 & \left.\vphantom{\frac{\boldsymbol{B}^{2}}{8\pi}}+\left(\frac{\boldsymbol{E}^{2}+\boldsymbol{B}^{2}}{8\pi}\right)\boldsymbol{I}-\frac{\boldsymbol{E}\boldsymbol{E}+\boldsymbol{B}\boldsymbol{B}}{4\pi}+\boldsymbol{B}\boldsymbol{M}_{0}-\left(\boldsymbol{M}_{0}\cdot\boldsymbol{B}\right)\boldsymbol{I}\right\} =0.\label{eq:113}
\end{align}
Substituting the polarization vector field $\boldsymbol{P}$ and magnetization
vector field $\boldsymbol{M}$ of the drift kinetic system (see Eq.$\thinspace$(\ref{eq:reduced_polarization_magnetization}))
into Eq.$\thinspace$(\ref{eq:113}), we have
\begin{align}
 & \frac{D}{Dt}\left\{ \sum_{a}m_{a}u_{a}\boldsymbol{b}\delta_{a}+\frac{\boldsymbol{E}\times\boldsymbol{B}}{4\pi c}\right\} +\frac{D}{D\boldsymbol{x}}\cdot\left\{ \sum_{a}m_{a}u_{a}^{2}\boldsymbol{b}+\sum_{a}m_{a}u_{a}\left(\dot{\boldsymbol{X}}_{a\perp}\boldsymbol{b}+\boldsymbol{b}\dot{\boldsymbol{X}}_{a\perp}\right)\delta_{a}\right.\nonumber \\
 & \left.\vphantom{\frac{\boldsymbol{B}^{2}}{8\pi}}\sum_{a}\mu_{a}B\delta_{a}\left(\boldsymbol{I}-\boldsymbol{bb}\right)+\left(\frac{\boldsymbol{E}^{2}+\boldsymbol{B}^{2}}{8\pi}\right)\boldsymbol{I}-\frac{\boldsymbol{E}\boldsymbol{E}+\boldsymbol{B}\boldsymbol{B}}{4\pi}\right\} =0.\label{eq:114}
\end{align}
In terms of the distribution function $F_{s}(t,\boldsymbol{x},\boldsymbol{u})$
and the electromagnetic field $\left(\boldsymbol{E}(t,\boldsymbol{x}),\boldsymbol{B}(t,\boldsymbol{x})\right),$
Eq.\,(\ref{eq:114}) is
\begin{align}
 & \frac{D}{Dt}\left\{ \sum_{s}m_{s}\int F_{s}u_{\parallel}\boldsymbol{b}d^{3}\boldsymbol{u}+\frac{\boldsymbol{E}\times\boldsymbol{B}}{4\pi c}\right\} +\frac{D}{D\boldsymbol{x}}\cdot\left\{ \sum_{s}\int F_{s}\left[m_{s}u_{\parallel}^{2}\boldsymbol{b}\boldsymbol{b}+m_{s}u_{\parallel}\left(\dot{\boldsymbol{X}}_{s\perp}\boldsymbol{b}+\boldsymbol{b}\dot{\boldsymbol{X}}_{s\perp}\right)\right.\right.\nonumber \\
 & \left.\vphantom{\frac{\boldsymbol{E}^{2}+\boldsymbol{B}^{2}}{8\pi}}\left.\vphantom{\dot{\boldsymbol{X}}_{s\perp}}\mu B\left(\boldsymbol{I}-\boldsymbol{bb}\right)\right]d^{3}\boldsymbol{u}+\left(\frac{\boldsymbol{E}^{2}+\boldsymbol{B}^{2}}{8\pi}\right)\boldsymbol{I}-\frac{\boldsymbol{E}\boldsymbol{E}+\boldsymbol{B}\boldsymbol{B}}{4\pi}\right\} =0.\label{eq:Brizard_momen}
\end{align}
Equation (\ref{eq:Brizard_momen}), as a special case of the gyrokinetic
momentum conservation law (\ref{eq:momentum_conservation-1}), is
consistent with the result shown by Brizard et al. \citep{Brizard2016a}
for the drift kinetics.

This completes our derivation and discussion of the momentum conservation
law for the first-order theory. 

For the general electromagnetic gyrokinetic system defined by Eq.\,(\ref{eq:General-GK-Lag}),
the following exact, gauge-invariant, local momentum conservation
law can be derived using a similar method,
\begin{align}
 & \frac{D}{Dt}\left\{ \sum_{s}\int d^{3}\boldsymbol{u}F_{s}\left[m_{s}\left(u_{\parallel}\boldsymbol{b}+\boldsymbol{D}\right)+\frac{1}{c}\boldsymbol{\mathfrak{p}}_{0s}\times\boldsymbol{B}\right]+\frac{\boldsymbol{E}\times\boldsymbol{B}}{4\pi c}+\sum_{s}\int d^{3}\boldsymbol{u}F_{s}\left[\frac{\partial\delta L_{s}}{\partial\dot{\boldsymbol{X}}_{s}}+\frac{1}{c}\left(\delta\mathfrak{\boldsymbol{p}}_{s}\times\boldsymbol{B}\right)\right.\right.\nonumber \\
 & \left.\vphantom{\frac{\partial L_{1s}}{\partial\dot{\boldsymbol{X}}_{s}}}\left.\vphantom{\frac{\partial L_{1s}}{\partial\dot{\boldsymbol{X}}_{s}}}+\delta\boldsymbol{K}\right]\right\} +\frac{D}{D\boldsymbol{x}}\cdot\left\{ \vphantom{\frac{\partial L_{1s}}{\partial\dot{\boldsymbol{X}}_{s}}}\sum_{s}\int d^{3}\boldsymbol{u}F_{s}\left[m_{s}\dot{\boldsymbol{X}}_{s}\left(u_{\parallel}\boldsymbol{b}+\boldsymbol{D}\right)\vphantom{\dot{\boldsymbol{X}}_{s}}\right.\right.\nonumber \\
 & \left.\vphantom{\dot{\boldsymbol{X}}_{s}}+\boldsymbol{B}\mathfrak{\boldsymbol{m}}_{0s}-\left(\boldsymbol{\mathfrak{m}}_{0s}\cdot\boldsymbol{B}\right)\boldsymbol{I}-\mathfrak{\boldsymbol{p}}_{0s}\boldsymbol{E}\right]+\left(\frac{\boldsymbol{E}^{2}+\boldsymbol{B}^{2}}{8\pi}\right)\boldsymbol{I}-\frac{\boldsymbol{E}\boldsymbol{E}+\boldsymbol{B}\boldsymbol{B}}{4\pi}+\nonumber \\
 & \left.\vphantom{\frac{\partial L_{1s}}{\partial\dot{\boldsymbol{X}}_{s}}}+\sum_{s}\int d^{3}\boldsymbol{u}F_{s}\left[\dot{\boldsymbol{X}}_{s}\frac{\partial\delta L_{s}}{\partial\dot{\boldsymbol{X}}_{S}}-\delta\mathfrak{\boldsymbol{p}}_{s}\boldsymbol{E}+\boldsymbol{B}\delta\mathfrak{\boldsymbol{m}}_{s}-\left(\boldsymbol{B}\cdot\delta\mathfrak{\boldsymbol{m}}_{s}\right)\boldsymbol{I}+\delta\mathcal{\boldsymbol{K}}\right]\right\} =0,\label{eq:general-GK-momentum-conser}
\end{align}
where $\delta\mathfrak{\boldsymbol{p}}_{s}$ and $\delta\mathfrak{\boldsymbol{m}}_{s}$
are defined in Eqs.\,(\ref{eq:delta-ps}) and (\ref{eq:delta-ms}),
and
\begin{align}
 & \delta\boldsymbol{K}=\sum_{i=1}^{n}\sum_{j=1}^{i}\left(-1\right)^{j}\left[D_{\mu_{1}}\cdots D_{\mu_{j-1}}\frac{\partial\delta L_{s}}{\partial D_{\mu_{1}}\cdots D_{\mu_{j-1}}\partial_{t}D_{\mu_{j+1}}\cdots D_{\mu_{i}}\boldsymbol{E}}\right]\cdot\left[D_{\mu_{j+1}}\cdots D_{\mu_{i}}\left(\boldsymbol{\nabla}\boldsymbol{E}\right)^{T}\right]\nonumber \\
 & +\sum_{i=1}^{n}\sum_{j=1}^{i}\left(-1\right)^{j}\left[D_{\mu_{1}}\cdots D_{\mu_{j-1}}\frac{\partial\delta L_{s}}{\partial D_{\mu_{1}}\cdots D_{\mu_{j-1}}\partial_{t}D_{\mu_{j+1}}\cdots D_{\mu_{i}}\boldsymbol{B}}\right]\cdot\left[D_{\mu_{j+1}}\cdots D_{\mu_{i}}\left(\boldsymbol{\nabla}\boldsymbol{B}\right)^{T}\right],\label{eq:delta-K}\\
 & \delta\mathcal{\boldsymbol{K}}=\sum_{i=1}^{n}\sum_{j=1}^{i}\left(-1\right)^{j}\left[D_{\mu_{1}}\cdots D_{\mu_{j-1}}\frac{\partial\delta L_{s}}{\partial D_{\mu_{1}}\cdots D_{\mu_{j-1}}\boldsymbol{\nabla}D_{\mu_{j+1}}\cdots D_{\mu_{i}}\boldsymbol{E}}\right]\cdot\left[D_{\mu_{j+1}}\cdots D_{\mu_{i}}\left(\boldsymbol{\nabla}\boldsymbol{E}\right)^{T}\right]\nonumber \\
 & +\sum_{i=1}^{n}\sum_{j=1}^{i}\left(-1\right)^{j}\left[D_{\mu_{1}}\cdots D_{\mu_{j-1}}\frac{\partial\delta L_{s}}{\partial D_{\mu_{1}}\cdots D_{\mu_{j-1}}\boldsymbol{\nabla}D_{\mu_{j+1}}\cdots D_{\mu_{i}}\boldsymbol{B}}\right]\cdot\left[D_{\mu_{j+1}}\cdots D_{\mu_{i}}\left(\boldsymbol{\nabla}\boldsymbol{B}\right)^{T}\right].\label{eq:delta-math-K}
\end{align}

\section{conclusion}

We have established the exact, gauge-invariant, local energy-momentum
conservation laws for the electromagnetic gyrokinetic system from
the underpinning spacetime translation symmetries of the system. Because
the gyrocenter and electromagnetic field are defined on different
manifolds, the standard Noether procedure for deriving conservation
laws from symmetries does not apply to the gyrokinetic system without
modification. 

To establish the connection between energy-momentum conservation and
spacetime translation symmetry for the electromagnetic gyrokinetic
system, we first extended the field theory for classical particle-field
system on heterogeneous manifolds $\thinspace$\citep{Qin2014b,Fan2018,Fan2019}
to include high-order field derivatives and using noncanonical phase
space coordinates in a general setting without specializing to the
gyrokinetic system. The field theory on heterogeneous manifolds embraces
the fact that for classical particle-field systems, particles and
fields reside on different manifolds, and a weak Euler-Lagrange equation
was developed to replace the standard Euler-Lagrange equation for
particles. The weak Euler-Lagrange current, induced by the weak Euler-Lagrange
equation, is the new physics associated with the field theory on heterogeneous
manifolds, and it plays a crucial role in the connection between symmetries
and conservation laws when different components of the system are
defined on different manifolds. 

The high-order field theory on heterogeneous manifolds developed was
then applied to the electromagnetic gyrokinetic system to derive the
exact, local energy-momentum conservation laws from the spacetime
translation symmetries admitted by the Lagrangian density of the system.
And, finally, the recently developed gauge-symmetrization procedure
\citep{Fan2021} using the electromagnetic displacement-potential
tensor was applied to render the conservation laws electromagnetic
gauge invariant.
\begin{acknowledgments}
P. Fan was supported by the Chinese Scholarship Council (CSC) with
No.$\thinspace$201806340074, Shenzhen Clean Energy Research Institute
and National Natural Science Foundation of China (NSFC-12005141).
H. Qin was supported by the U.S. Department of Energy (DE-AC02-09CH11466).
J. Xiao was supported by the National MC Energy R\&D Program (2018YFE0304100),
National Key Research and Development Program (2016YFA0400600, 2016YFA0400601
and 2016YFA0400602), and the National Natural Science Foundation of
China (NSFC-11905220 and 11805273). 
\end{acknowledgments}

\appendix

\section{Derivations of polarization and magnetization in Eqs.$\thinspace$
(\ref{eq:polarization}) and (\ref{eq:magnetization}) \label{sec:polarization and magnetization}}

In this appendix, we give the derivations of zeroth order polorization
$\boldsymbol{P}_{0}$ and magnetization $\boldsymbol{M}_{0}$. From
the definition of $\boldsymbol{P}_{0}$, $\boldsymbol{M}_{0}$ and
Lagrangian density of the a-th particle (see Eqs.$\thinspace$ (\ref{eq:polarization}),
(\ref{eq:magnetization}), (\ref{eq:1st-GK-Lag}) and (\ref{eq:particle_lagrangian_density})),
they are derived as follows
\begin{align}
 & \boldsymbol{P}_{0}=\sum_{a}\frac{\partial\mathcal{L}_{0a}}{\partial\boldsymbol{E}}\nonumber \\
 & =\sum_{a}\left\{ \frac{\partial}{\partial\boldsymbol{E}}\left[\frac{m_{a}c\delta_{a}}{B}\boldsymbol{E}\cdot\left(\boldsymbol{b}\times\dot{\boldsymbol{X}}_{a}\right)\right]-\frac{\partial}{\partial\boldsymbol{E}}\left(\frac{1}{2}m_{a}\boldsymbol{D}^{2}\delta_{a}\right)\right\} \nonumber \\
 & =\sum_{a}\left\{ \left[\frac{m_{a}c\delta_{a}}{B}\frac{\partial\boldsymbol{E}}{\partial\boldsymbol{E}}\cdot\left(\boldsymbol{b}\times\dot{\boldsymbol{X}}_{a}\right)\right]-m_{a}\delta_{a}\frac{\partial\boldsymbol{D}}{\partial\boldsymbol{E}}\cdot\boldsymbol{D}\right\} \nonumber \\
 & =\sum_{a}\left\{ \left[\frac{m_{a}c\delta_{a}}{B}\boldsymbol{I}\cdot\left(\boldsymbol{b}\times\dot{\boldsymbol{X}}_{a}\right)\right]-m_{a}c\delta_{a}\left[\boldsymbol{I}\times\left(\frac{\boldsymbol{b}}{B}\right)\right]\cdot\boldsymbol{D}\right\} \nonumber \\
 & =\sum_{a}\left\{ \left[\frac{m_{a}c\delta_{a}}{B}\left(\boldsymbol{b}\times\dot{\boldsymbol{X}}_{a}\right)\right]-m_{a}c\delta_{a}\left[\left(\frac{\boldsymbol{b}}{B}\right)\times\boldsymbol{D}\right]\right\} \nonumber \\
 & =\sum_{a}\frac{m_{a}c\delta_{a}}{B}\left[\boldsymbol{b}\times\left(\dot{\boldsymbol{X}}_{a}-\boldsymbol{D}\right)\right]\label{eq:116}
\end{align}
and
\begin{align}
 & \boldsymbol{M}_{0}=\sum_{a}\frac{\partial\mathcal{L}_{0a}}{\partial\boldsymbol{B}}\nonumber \\
 & =\sum_{a}\frac{\partial}{\partial\boldsymbol{B}}\left[m_{a}u_{a}\delta_{a}\boldsymbol{b}\cdot\dot{\boldsymbol{X}}_{a}+m_{a}\delta_{a}\boldsymbol{D}\cdot\dot{\boldsymbol{X}}_{a}-\frac{1}{2}m_{a}\delta_{a}\boldsymbol{D}^{2}-\mu_{a}\delta_{a}B\right]\nonumber \\
 & =\sum_{a}\left[m_{a}u_{a}\delta_{a}\frac{\partial\boldsymbol{b}}{\partial\boldsymbol{B}}\cdot\dot{\boldsymbol{X}}_{a}+m_{a}\delta_{a}\frac{\partial\boldsymbol{D}}{\partial\boldsymbol{B}}\cdot\dot{\boldsymbol{X}}_{a}-\frac{1}{2}m_{a}\delta_{a}\frac{\partial\boldsymbol{D}^{2}}{\partial\boldsymbol{B}}-\mu_{a}\delta_{a}\frac{\partial B}{\partial\boldsymbol{B}}\right]\nonumber \\
 & =\sum_{a}\left[m_{a}u_{a}\delta_{a}\left(\frac{\boldsymbol{I}-\boldsymbol{b}\boldsymbol{b}}{B}\right)\cdot\dot{\boldsymbol{X}}_{a}-\frac{m_{a}c\delta_{a}}{B^{2}}\left[\left(\boldsymbol{I}-2\boldsymbol{b}\boldsymbol{b}\right)\times\boldsymbol{E}\right]\cdot\left(\dot{\boldsymbol{X}}_{a}-\boldsymbol{D}\right)-\mu_{a}\delta_{a}\boldsymbol{b}\right]\nonumber \\
 & =\sum_{a}\left[m_{a}u_{a}\dot{\boldsymbol{X}}_{a\perp}-\mu_{a}\delta_{a}\boldsymbol{b}+\left[-\frac{m_{a}c\delta_{a}}{B^{2}}\left(\boldsymbol{I}\times\boldsymbol{E}\right)-\frac{2m_{a}\delta_{a}}{B}\boldsymbol{b}\boldsymbol{D}\right]\cdot\left(\dot{\boldsymbol{X}}_{a}-\boldsymbol{D}\right)\right]\nonumber \\
 & =\sum_{a}\frac{m_{a}c\delta_{a}}{B}\left[\frac{u_{a}}{c}\dot{\boldsymbol{X}}_{a\perp}-\frac{\mu_{a}B}{m_{a}c}\boldsymbol{b}-\frac{\boldsymbol{E}}{B}\times\left(\dot{\boldsymbol{X}}_{a}-\boldsymbol{D}\right)-\frac{2}{c}\left[\left(\dot{\boldsymbol{X}}_{a}-\boldsymbol{D}\right)\cdot\boldsymbol{D}\right]\boldsymbol{b}\right].\label{eq:117}
\end{align}
In obtaining Eqs.$\thinspace$(\ref{eq:116}) and (\ref{eq:117}),
the following equations were used
\begin{align}
 & \frac{\partial\boldsymbol{D}}{\partial\boldsymbol{E}}=c\frac{\partial}{\partial\boldsymbol{E}}\left(\frac{\boldsymbol{E}\times\boldsymbol{B}}{B^{2}}\right)=c\frac{\partial\boldsymbol{E}}{\partial\boldsymbol{E}}\times\left(\frac{\boldsymbol{b}}{B}\right)=c\boldsymbol{I}\times\left(\frac{\boldsymbol{b}}{B}\right),\label{eq:118}\\
 & \frac{\partial B}{\partial\boldsymbol{B}}=\frac{\partial\sqrt{\boldsymbol{B}^{2}}}{\partial\boldsymbol{B}}=\frac{1}{\sqrt{\boldsymbol{B}^{2}}}\frac{1}{2}\frac{\partial\boldsymbol{B}^{2}}{\partial\boldsymbol{B}}=\frac{\boldsymbol{B}}{B}=\boldsymbol{b},\label{eq:119}\\
 & \frac{\partial\boldsymbol{b}}{\partial\boldsymbol{B}}=\frac{\partial}{\partial\boldsymbol{B}}\left(\frac{\boldsymbol{B}}{B}\right)=\left[\frac{1}{B}\frac{\partial\boldsymbol{B}}{\partial\boldsymbol{B}}+\frac{\partial}{\partial\boldsymbol{B}}\left(\frac{1}{B}\right)\boldsymbol{B}\right]=\left[\frac{\boldsymbol{I}}{B}-\frac{1}{B^{2}}\frac{\partial B}{\partial\boldsymbol{B}}\boldsymbol{B}\right]=\frac{\boldsymbol{I}-\boldsymbol{b}\boldsymbol{b}}{B},\label{eq:120}\\
 & \frac{\partial\boldsymbol{D}}{\partial\boldsymbol{B}}=-c\frac{\partial}{\partial\boldsymbol{B}}\left(\frac{\boldsymbol{B}\times\boldsymbol{E}}{B^{2}}\right)=-c\frac{\partial}{\partial\boldsymbol{B}}\left(\frac{\boldsymbol{B}}{B^{2}}\right)\times\boldsymbol{E}\nonumber \\
 & =-c\left\{ \left[\frac{\partial}{\partial\boldsymbol{B}}\left(\frac{1}{B^{2}}\right)\boldsymbol{B}+\frac{1}{B^{2}}\frac{\partial\boldsymbol{B}}{\partial\boldsymbol{B}}\right]\times\boldsymbol{E}\right\} =-\frac{c}{B^{2}}\left(\boldsymbol{I}-2\boldsymbol{b}\boldsymbol{b}\right)\times\boldsymbol{E}.\label{eq:121}
\end{align}

\section{Derivations of Eqs.$\thinspace$(\ref{eq:74})-(\ref{eq:77}) and
Eqs.$\thinspace$(\ref{eq:98})-(\ref{eq:101})\label{sec:Boundary-terms}}

In this appendix, we show the detailed derivations of Eqs.$\thinspace$(\ref{eq:74})-(\ref{eq:77})
and Eqs.$\thinspace$(\ref{eq:98})-(\ref{eq:101}), which are boundary
terms induced by time and space translation symmetries. For the time
translation symmetry, using Eqs.$\thinspace$(\ref{eq:24}) and (\ref{eq:26})-(\ref{eq:30-1}),
equations (\ref{eq:74})-(\ref{eq:77}) can be proved as follows

\begin{align}
 & \sum_{a}\mathscr{P}_{a(1)}^{\nu}\delta_{a}+\mathbb{P}_{F\left(1\right)}^{\nu}=Q^{\alpha}\frac{\partial\mathcal{L}_{0}}{\partial\left(\partial_{\nu}\psi^{\alpha}\right)}+Q^{\alpha}\frac{\partial\mathcal{L}_{1}}{\partial\left(\partial_{\nu}\psi^{\alpha}\right)}\nonumber \\
 & =\left(-\frac{\partial\mathcal{L}_{0}}{\partial\left(\partial_{t}\varphi\right)}\varphi_{,t}-\frac{\partial\mathcal{L}_{0}}{\partial\left(\partial_{t}\boldsymbol{A}\right)}\cdot\boldsymbol{A}_{,t},-\frac{\partial\mathcal{L}_{0}}{\partial\left(\boldsymbol{\nabla}\varphi\right)}\varphi_{,t}-\frac{\partial\mathcal{L}_{0}}{\partial\left(\boldsymbol{\nabla}\boldsymbol{A}\right)}\cdot\boldsymbol{A}_{,t}\right)+\sum_{a}\mathscr{P}_{1a\left(1\right)}^{\nu}\delta_{a}\nonumber \\
 & =\frac{1}{4\pi}\left(\frac{1}{c}\left(\boldsymbol{E}+4\pi\boldsymbol{P}_{0}\right)\cdot\boldsymbol{A}_{,t},\left(\boldsymbol{E}+4\pi\boldsymbol{P}_{0}\right)\varphi_{,t}+\boldsymbol{A}_{,t}\times\left(\boldsymbol{B}-4\pi\boldsymbol{M}_{0}\right)\right)+\sum_{a}\mathscr{P}_{1a\left(1\right)}^{\nu}\delta_{a},\label{eq:122}\\
 & \mathscr{P}_{1a\left(1\right)}^{\nu}=Q^{\alpha}\frac{\partial L_{1a}}{\partial\left(\partial_{\nu}\psi^{\alpha}\right)}=\left(-\frac{\partial L_{1a}}{\partial\left(\partial_{t}\boldsymbol{A}\right)}\cdot\boldsymbol{A}_{,t},-\frac{\partial L_{1a}}{\partial\left(\boldsymbol{\nabla}\varphi\right)}\varphi_{,t}-\frac{\partial L_{1a}}{\partial\left(\boldsymbol{\nabla}\boldsymbol{A}\right)}\cdot\boldsymbol{A}_{,t}\right)\nonumber \\
 & =\left(\frac{1}{c}\frac{\partial L_{1a}}{\partial\boldsymbol{E}}\cdot\boldsymbol{A}_{,t},\;\frac{\partial L_{1a}}{\partial\boldsymbol{E}}\varphi_{,t}-\boldsymbol{A}_{,t}\times\frac{\partial L_{1a}}{\partial\boldsymbol{B}}\right),\label{eq:123}\\
 & \mathbb{P}_{F\left(2\right)}^{\nu}=D_{\mu}Q^{\alpha}\left[\frac{\partial\mathcal{L}_{F}}{\partial\left(\partial_{\nu}\partial_{\mu}\psi^{\alpha}\right)}\right]-Q^{\alpha}\left[D_{\mu}\frac{\partial\mathcal{L}_{F}}{\partial\left(\partial_{\mu}\partial_{\nu}\psi^{\alpha}\right)}\right]=0,\label{eq:124}\\
 & \mathscr{P}_{a\left(2\right)}^{\nu}=D_{\mu}Q^{\alpha}\left[\frac{\partial L_{a}}{\partial\left(\partial_{\nu}\partial_{\mu}\psi^{\alpha}\right)}\right]-Q^{\alpha}\left[D_{\mu}\frac{\partial L_{a}}{\partial\left(\partial_{\mu}\partial_{\nu}\psi^{\alpha}\right)}\right]\nonumber \\
 & =\left(-\left[\frac{\partial L_{1a}}{\partial\left(\partial_{t}\varphi_{,t}\right)}\right]\partial_{t}\varphi_{,t}-\left[\frac{\partial L_{1a}}{\partial\left(\partial_{t}\boldsymbol{\nabla}\varphi\right)}\right]\cdot\partial_{t}\boldsymbol{\nabla}\varphi-\left[\frac{\partial L_{1a}}{\partial\left(\partial_{t}\boldsymbol{A}_{,t}\right)}\right]\cdot\partial_{t}\boldsymbol{A}_{,t}\right.\nonumber \\
 & -\left[\frac{\partial L_{1a}}{\partial\left(\partial_{t}\boldsymbol{\nabla}\boldsymbol{A}\right)}\right]:\partial_{t}\boldsymbol{\nabla}\boldsymbol{A}+\left[D_{\mu}\frac{\partial L_{1a}}{\partial\left(\partial_{\mu}\partial_{t}\varphi\right)}\right]\varphi_{,t}+\left[D_{\mu}\frac{\partial L_{1a}}{\partial\left(\partial_{\mu}\partial_{t}\boldsymbol{A}\right)}\right]\cdot\boldsymbol{A}_{,t},\nonumber \\
 & -\left[\frac{\partial L_{1a}}{\partial\left(\boldsymbol{\nabla}\partial_{t}\varphi\right)}\right]\partial_{t}\varphi_{,t}-\left[\frac{\partial L_{1a}}{\partial\left(\boldsymbol{\nabla}\boldsymbol{\nabla}\varphi\right)}\right]\cdot\boldsymbol{\nabla}\varphi_{,t}-\left[\frac{\partial L_{1a}}{\partial\left(\boldsymbol{\nabla}\partial_{t}\boldsymbol{A}\right)}\right]\cdot\partial_{t}\boldsymbol{A}_{,t}-\left[\frac{\partial L_{1a}}{\partial\left(\boldsymbol{\nabla}\boldsymbol{\nabla}\boldsymbol{A}\right)}\right]:\boldsymbol{\nabla}\boldsymbol{A}_{,t}\nonumber \\
 & \left.\vphantom{\left[\frac{\partial\mathcal{L}_{1}}{\partial\left(\partial_{\mu}\right)}\right]}+\left[D_{\mu}\frac{\partial L_{1a}}{\partial\left(\partial_{\mu}\boldsymbol{\nabla}\varphi\right)}\right]\varphi_{,t}+\left[D_{\mu}\frac{\partial L_{1a}}{\partial\left(\partial_{\mu}\boldsymbol{\nabla}\boldsymbol{A}\right)}\right]\cdot\boldsymbol{A}_{,t}\right)\nonumber \\
 & =\left(-\left[\frac{\partial L_{1a}}{\partial\left(\partial_{t}\boldsymbol{E}\right)}\right]\cdot\partial_{t}\boldsymbol{E}-\left[\frac{\partial L_{1a}}{\partial\left(\partial_{t}\boldsymbol{B}\right)}\right]\cdot\partial_{t}\boldsymbol{B}-\frac{1}{c}\left[D_{\mu}\frac{\partial L_{1a}}{\partial\left(\partial_{\mu}\boldsymbol{E}\right)}\right]\cdot\boldsymbol{A}_{,t},\right.\nonumber \\
 & \left.\vphantom{\left[\frac{\partial\mathcal{L}_{1}}{\partial\left(\partial_{\mu}\right)}\right]}-\left[\frac{\partial L_{1a}}{\partial\left(\boldsymbol{\nabla}\boldsymbol{E}\right)}\right]\cdot\partial_{t}\boldsymbol{E}-\left[\frac{\partial L_{1a}}{\partial\left(\boldsymbol{\nabla}\boldsymbol{B}\right)}\right]\cdot\partial_{t}\boldsymbol{B}-\left[D_{\mu}\frac{\partial L_{1a}}{\partial\left(\partial_{\mu}\boldsymbol{E}\right)}\right]\varphi_{,t}+\boldsymbol{A}_{,t}\times\left[D_{\mu}\frac{\partial L_{1a}}{\partial\left(\partial_{\mu}\boldsymbol{B}\right)}\right]\right).\label{eq:125}
\end{align}
Similarly, for the space translation symmetry, using the definitions
of $\mathscr{P}_{a}^{\nu}$ and $\mathbb{P}_{F}^{v}$ (see Eqs.$\thinspace$(\ref{eq:26})-(\ref{eq:30-1})),
equations (\ref{eq:98})-(\ref{eq:101}) are demonstrated as follows
\begin{align}
 & \sum_{a}\mathscr{P}_{a(1)}^{\nu}\delta_{a}+\mathbb{P}_{F\left(1\right)}^{\nu}\nonumber \\
 & =Q^{\alpha}\frac{\partial\mathcal{L}_{0}}{\partial\left(\partial_{\nu}\psi^{\alpha}\right)}+Q^{\alpha}\frac{\partial\mathcal{L}_{1}}{\partial\left(\partial_{\nu}\psi^{\alpha}\right)}\nonumber \\
 & =\left(-\frac{\partial\mathcal{L}_{0}}{\partial\left(\partial_{t}\varphi\right)}\boldsymbol{\nabla}\varphi-\frac{\partial\mathcal{L}_{0}}{\partial\left(\partial_{t}\boldsymbol{A}\right)}\cdot\boldsymbol{\nabla}\boldsymbol{A},-\frac{\partial\mathcal{L}_{0}}{\partial\left(\boldsymbol{\nabla}\varphi\right)}\boldsymbol{\nabla}\varphi-\frac{\partial\mathcal{L}_{0}}{\partial\left(\boldsymbol{\nabla}\boldsymbol{A}\right)}\cdot\left(\boldsymbol{\nabla}\boldsymbol{A}\right)^{T}\right)\cdot\boldsymbol{h}\nonumber \\
 & +Q^{\alpha}\frac{\partial\mathcal{L}_{1}}{\partial\left(\partial_{\nu}\psi^{\alpha}\right)}\nonumber \\
 & =\frac{1}{4\pi}\left(\frac{1}{c}\left(\boldsymbol{E}+4\pi\boldsymbol{P}_{0}\right)\cdot\left(\boldsymbol{\nabla}\boldsymbol{A}\right)^{T},\left(\boldsymbol{E}+4\pi\boldsymbol{P}_{0}\right)\boldsymbol{\nabla}\varphi-\boldsymbol{\varepsilon}:\left[\left(\boldsymbol{B}-4\pi\boldsymbol{M}_{0}\right)\left(\boldsymbol{\nabla}\boldsymbol{A}\right)^{T}\right]\right)\cdot\boldsymbol{h}\nonumber \\
 & +\left(\sum_{a}\boldsymbol{\sigma}_{1a\left(1\right)}^{\nu}\delta_{a}\right)\cdot\boldsymbol{h},\label{eq:126}\\
 & \boldsymbol{\sigma}_{1a\left(1\right)}^{\nu}=-\frac{\partial L_{1a}}{\partial\left(\partial_{\nu}\psi^{\alpha}\right)}\boldsymbol{\nabla}\psi^{\alpha}=\left(\boldsymbol{\sigma}_{1a\left(1\right)}^{0},\boldsymbol{\sigma}_{1a\left(1\right)}\right)\nonumber \\
 & =\left(-\frac{\partial L_{1a}}{\partial\left(\partial_{t}\varphi\right)}\boldsymbol{\nabla}\varphi-\frac{\partial L_{1a}}{\partial\left(\partial_{t}\boldsymbol{A}\right)}\cdot\left(\boldsymbol{\nabla}\boldsymbol{A}\right)^{T},-\frac{\partial L_{1a}}{\partial\left(\boldsymbol{\nabla}\varphi\right)}\boldsymbol{\nabla}\varphi-\frac{\partial L_{1a}}{\partial\left(\boldsymbol{\nabla}\boldsymbol{A}\right)}\cdot\left(\boldsymbol{\nabla}\boldsymbol{A}\right)^{T}\right)\nonumber \\
 & =\left(\frac{1}{c}\frac{\partial L_{1a}}{\partial\boldsymbol{E}}\cdot\left(\boldsymbol{\nabla}\boldsymbol{A}\right)^{T},\frac{\partial L_{1a}}{\partial\boldsymbol{E}}\boldsymbol{\nabla}\varphi+\frac{\partial L_{1a}}{\partial\boldsymbol{B}}\times\left(\boldsymbol{\nabla}\boldsymbol{A}\right)^{T}\right),\label{eq:127}\\
 & \sum_{a}\mathscr{P}_{a(2)}^{\nu}\delta_{a}+\mathbb{P}_{F\left(2\right)}^{\nu}=D_{\mu}Q^{\alpha}\left[\frac{\partial\mathcal{L}}{\partial\left(\partial_{\nu}\partial_{\mu}\psi^{\alpha}\right)}\right]-Q^{\alpha}\left[D_{\mu}\frac{\partial\mathcal{L}}{\partial\left(\partial_{\mu}\partial_{\nu}\psi^{\alpha}\right)}\right]=\left(\sum_{a}\boldsymbol{\sigma}_{a\left(2\right)}^{\nu}\delta_{a}\right)\cdot\boldsymbol{h},\label{eq:128}\\
 & \boldsymbol{\sigma}_{a\left(2\right)}^{\nu}=-\left[\frac{\partial L_{1a}}{\partial\left(\partial_{\nu}\partial_{\mu}\psi^{\alpha}\right)}\right]D_{\mu}\left(\boldsymbol{\nabla}\psi^{\alpha}\right)+\left[D_{\mu}\frac{\partial L_{1a}}{\partial\left(\partial_{\mu}\partial_{\nu}\psi^{\alpha}\right)}\right]\left(\boldsymbol{\nabla}\psi^{\alpha}\right)=\left(\boldsymbol{\sigma}_{a\left(2\right)}^{0},\boldsymbol{\sigma}_{a\left(2\right)}\right)\nonumber \\
 & =\left(-\left[\frac{\partial L_{1a}}{\partial\left(\partial_{t}\partial_{t}\varphi\right)}\right]\partial_{t}\left(\boldsymbol{\nabla}\varphi\right)-\left[\frac{\partial L_{1a}}{\partial\left(\partial_{t}\boldsymbol{\nabla}\varphi\right)}\right]\cdot\boldsymbol{\nabla}\left(\boldsymbol{\nabla}\varphi\right)-\left[\frac{\partial L_{1a}}{\partial\left(\partial_{t}\partial_{t}\boldsymbol{A}\right)}\right]\cdot\partial_{t}\left(\boldsymbol{\nabla}\boldsymbol{A}\right)^{T}\right.\nonumber \\
 & -\left[\frac{\partial L_{1a}}{\partial\left(\partial_{t}\boldsymbol{\nabla}\boldsymbol{A}\right)}\right]:\boldsymbol{\nabla}\left(\boldsymbol{\nabla}\boldsymbol{A}\right)^{T}+\left[D_{\mu}\frac{\partial L_{1a}}{\partial\left(\partial_{\mu}\partial_{t}\varphi\right)}\right]\left(\boldsymbol{\nabla}\varphi\right)+\left[D_{\mu}\frac{\partial L_{1a}}{\partial\left(\partial_{\mu}\partial_{t}\boldsymbol{A}\right)}\right]\cdot\left(\boldsymbol{\nabla}\boldsymbol{A}\right)^{T},\nonumber \\
 & -\left[\frac{\partial L_{1a}}{\partial\left(\boldsymbol{\nabla}\partial_{t}\varphi\right)}\right]\partial_{t}\left(\boldsymbol{\nabla}\varphi\right)-\left[\frac{\partial L_{1a}}{\partial\left(\boldsymbol{\nabla}\boldsymbol{\nabla}\varphi\right)}\right]\cdot\boldsymbol{\nabla}\left(\boldsymbol{\nabla}\varphi\right)-\left[\frac{\partial L_{1a}}{\partial\left(\boldsymbol{\nabla}\partial_{t}\boldsymbol{A}\right)}\right]\cdot\partial_{t}\left(\boldsymbol{\nabla}\boldsymbol{A}\right)^{T}\nonumber \\
 & \left.\vphantom{\left[D_{\mu}\frac{\partial L_{1a}}{\partial\left(\partial_{\mu}\boldsymbol{\nabla}\boldsymbol{A}\right)}\right]}-\left[\frac{\partial L_{1a}}{\partial\left(\boldsymbol{\nabla}\boldsymbol{\nabla}\boldsymbol{A}\right)}\right]:\boldsymbol{\nabla}\left(\boldsymbol{\nabla}\boldsymbol{A}\right)^{T}+\left[D_{\mu}\frac{\partial L_{1a}}{\partial\left(\partial_{\mu}\boldsymbol{\nabla}\varphi\right)}\right]\left(\boldsymbol{\nabla}\varphi\right)+\left[D_{\mu}\frac{\partial L_{1a}}{\partial\left(\partial_{\mu}\boldsymbol{\nabla}\boldsymbol{A}\right)}\right]\cdot\left(\boldsymbol{\nabla}\boldsymbol{A}\right)^{T}\right)\nonumber \\
 & =\left(\left[\frac{\partial L_{1a}}{\partial\left(\partial_{t}\boldsymbol{E}\right)}\right]\cdot\boldsymbol{\nabla}\left(\boldsymbol{\nabla}\varphi\right)+\frac{1}{c}\left[\frac{\partial L_{1a}}{\partial\left(\partial_{t}\boldsymbol{E}\right)}\right]\cdot\partial_{t}\left(\boldsymbol{\nabla}\boldsymbol{A}\right)^{T}-\boldsymbol{\nabla}\boldsymbol{B}\cdot\left[\frac{\partial L_{1a}}{\partial\left(\partial_{t}\boldsymbol{B}\right)}\right]\right.\nonumber \\
 & -\left[\frac{1}{c}D_{\mu}\frac{\partial L_{1a}}{\partial\left(\partial_{\mu}\boldsymbol{E}\right)}\right]\cdot\left(\boldsymbol{\nabla}\boldsymbol{A}\right)^{T},\;\left[\frac{\partial L_{1a}}{\partial\left(\boldsymbol{\nabla}\boldsymbol{E}\right)}\right]\cdot\boldsymbol{\nabla}\left(\boldsymbol{\nabla}\varphi\right)\nonumber \\
 & +\frac{1}{c}\left[\frac{\partial L_{1a}}{\partial\left(\boldsymbol{\nabla}\boldsymbol{E}\right)}\right]\cdot\partial_{t}\left(\boldsymbol{\nabla}\boldsymbol{A}\right)^{T}-\left[\frac{\partial L_{1a}}{\partial\left(\boldsymbol{\nabla}\boldsymbol{B}\right)}\right]\cdot\left(\boldsymbol{\nabla}\boldsymbol{B}\right){}^{T}\nonumber \\
 & \left.\vphantom{\left[\frac{\partial L_{1a}}{\partial\left(\boldsymbol{\nabla}\boldsymbol{B}\right)}\right]}-\left[D_{\mu}\frac{\partial L_{1a}}{\partial\left(\partial_{\mu}\boldsymbol{E}\right)}\right]\left(\boldsymbol{\nabla}\varphi\right)-\left[D_{\mu}\frac{\partial L_{1a}}{\partial\left(\partial_{\mu}\boldsymbol{B}\right)}\right]\times\left(\boldsymbol{\nabla}\boldsymbol{A}\right)^{T}\right)\nonumber \\
 & =\left(-\boldsymbol{\nabla}\boldsymbol{E}\cdot\left[\frac{\partial L_{1a}}{\partial\left(\partial_{t}\boldsymbol{E}\right)}\right]-\boldsymbol{\nabla}\boldsymbol{B}\cdot\left[\frac{\partial L_{1a}}{\partial\left(\partial_{t}\boldsymbol{B}\right)}\right]-\left[\frac{1}{c}D_{\mu}\frac{\partial L_{1a}}{\partial\left(\partial_{\mu}\boldsymbol{E}\right)}\right]\cdot\left(\boldsymbol{\nabla}\boldsymbol{A}\right)^{T},\right.\nonumber \\
 & -\left[\frac{\partial L_{1a}}{\partial\left(\boldsymbol{\nabla}\boldsymbol{E}\right)}\right]\cdot\left(\boldsymbol{\nabla}\boldsymbol{E}\right)^{T}-\left[\frac{\partial L_{1a}}{\partial\left(\boldsymbol{\nabla}\boldsymbol{B}\right)}\right]\cdot\left(\boldsymbol{\nabla}\boldsymbol{B}\right){}^{T}\nonumber \\
 & \left.\vphantom{D_{\mu}\frac{\partial L_{1a}}{\partial\left(\partial_{\mu}\boldsymbol{B}\right)}}-\left[D_{\mu}\frac{\partial L_{1a}}{\partial\left(\partial_{\mu}\boldsymbol{E}\right)}\right]\left(\boldsymbol{\nabla}\varphi\right)-\left[D_{\mu}\frac{\partial L_{1a}}{\partial\left(\partial_{\mu}\boldsymbol{B}\right)}\right]\times\left(\boldsymbol{\nabla}\boldsymbol{A}\right)^{T}\right).\label{eq:129}
\end{align}

\bibliographystyle{apsrev4-1}
\bibliography{GKConservation}

\end{document}